\documentclass[twocolumn,pra, aps, floatfix,superscriptaddress]{revtex4}

\usepackage{extarrows}
\usepackage{lineno}
\usepackage{mathptmx}
\usepackage{subfigure}
\usepackage{dcolumn}
\usepackage{amsmath,amssymb}
\usepackage{bm}
\usepackage{color}
\usepackage{overpic}
\usepackage{latexsym}
\usepackage{epstopdf}
\usepackage{color}
\usepackage[english]{babel}
\usepackage{latexsym}
\usepackage{stmaryrd}
\usepackage{bigints}
\usepackage{psfrag,graphicx} 
\usepackage{epsf} 
\usepackage{subfigure} 
\usepackage{amsmath} 
\usepackage{amssymb} 
\usepackage{amsfonts}
\usepackage{bm}
\usepackage{natbib}
\usepackage{epstopdf}
\usepackage{appendix}
\usepackage{relsize}

  \usepackage{amsmath}
\usepackage{verbatim}
\usepackage{graphicx}
\usepackage{color}
\usepackage{tikz}

\definecolor{mygrey}{gray}{0.35}
\definecolor{myblue}{rgb}{0.2,0.2,0.8}
\definecolor{myzard}{cmyk}{0,0,0.05,0}
\definecolor{mywhite}{rgb}{1,1,1}
\definecolor{myred}{rgb}{1,0.,0.3}

\usepackage[colorlinks=true,citecolor=myblue,linkcolor=myred]{hyperref}
\def\be{\begin{equation}}
\def\ee{\end{equation}}
\def\ba{\begin{align}}
\def\enda{\end{align}}
\def\bi{\begin{itemize}}
\def\ei{\end{itemize}}
\def\bs#1{\boldsymbol{#1}}

 \def\ee{\mathord{\rm e}}
 
 \def\ii{\mathord{\rm i}}

 \def\ee{\mathord{\rm e}}
 
 \def\ii{\mathord{\rm i}}

\renewcommand{\ii}{{\rm i}}
\renewcommand{\ee}{{\rm e}}

\def\beq{\begin{equation}}
\def\beq{\begin{equation}}
\def\eeq{\end{equation}}

 \newcommand{\ket}[1]{|#1\rangle}
 \newcommand{\bra}[1]{\langle #1|}
 \newcommand{\braket}[2]{\langle #1|#2\rangle}

\begin{document}


\title{Symmetry-Breaking Topological Insulators in the $\mathbb{Z}_2$ Bose-Hubbard Model}

%
%

\author{Daniel Gonz\'{a}lez-Cuadra}\email{daniel.gonzalez@icfo.eu}
\affiliation{ICFO-Institut de Ci\`encies Fot\`oniques, The Barcelona Institute of Science and Technology, Av. Carl Friedrich Gauss 3, 08860 Barcelona, Spain}
\author{Alexandre Dauphin}\email{alexandre.dauphin@icfo.eu}
\affiliation{ICFO-Institut de Ci\`encies Fot\`oniques, The Barcelona Institute of Science and Technology, Av. Carl Friedrich Gauss 3, 08860 Barcelona, Spain}
\author{Przemys{\l}aw R. Grzybowski}
\affiliation{ICFO-Institut de Ci\`encies Fot\`oniques, The Barcelona Institute of Science and Technology, Av. Carl Friedrich Gauss 3, 08860 Barcelona, Spain}
\affiliation{Faculty of Physics, Adam Mickiewicz University, Umultowska 85, 61-614 Pozna{\'n}, Poland}
\author{Pawe\l~W\'ojcik}
\affiliation{ICFO-Institut de Ci\`encies Fot\`oniques, The Barcelona Institute of Science and Technology, Av. Carl Friedrich Gauss 3, 08860 Barcelona, Spain}
\affiliation{Faculty of Physics, University of Warsaw, Pasteura 5, 02-093 Warsaw, Poland}
\author{Maciej Lewenstein}
\affiliation{ICFO-Institut de Ci\`encies Fot\`oniques, The Barcelona Institute of Science and Technology, Av. Carl Friedrich Gauss 3, 08860 Barcelona, Spain}
\affiliation{ICREA-Instituci\'o Catalana de Recerca i Estudis Avan\c cats, Lluis Companys 23, 08010 Barcelona, Spain}
\author{Alejandro Bermudez}
\affiliation{Departamento de F\'{i}sica Te\'{o}rica, Universidad Complutense, 28040 Madrid, Spain}

\begin{abstract}

In this work, we study a one-dimensional model of interacting bosons coupled to a dynamical $\mathbb{Z}_2$ field, the $\mathbb{Z}_2$ Bose-Hubbard model, and analyze the interplay between spontaneous symmetry breaking and topological symmetry protection. In a previous work, we showed how this model exhibits a spontaneous breaking of the translational symmetry through a bosonic Peierls transition. Here we find how, at half filling, the resulting phase also displays topological features that coexist with the presence of long-range order and yields a \textit{topological bond order wave}. Using both analytical and numerical methods, we describe the properties of this phase,  showing that it cannot be adiabatically connected to a bosonic topological phase with vanishing Hubbard interactions, and thus constitutes an instance of an interaction-induced symmetry-breaking topological insulator. 

\end{abstract}

\maketitle
\setcounter{tocdepth}{2}
\begingroup
\hypersetup{linkcolor=black}
\tableofcontents
\endgroup

\section{Introduction}

{\it Emergence}~\cite{Anderson_more_is_diff} and {\it symmetry}~\cite{landau_symm_breaking} are primary driving forces for  the vast diversity of    phenomena   observed in condensed matter. On the one hand, emergence  can account for the appearance of different collective behaviour at macroscopic scales, starting from a collection of  many individual  particles that interact   quantum-mechanically (i.e.  quantum many-body effects) according to the same microscopic laws. On the other hand,  symmetry can explain how, sometimes, this complexity can   be tamed  by understanding the underlying microscopic symmetries, and how these are spontaneously broken to yield 
various phases of matter. According to Landau's  paradigm, even when the system Hamiltonian $H(g)$ commutes with a unitary operator $U_{\mathcal{G}}$  that describes a particular symmetry group $\mathcal{G}$, $U_{\mathcal{G}}H(g)=H(g)U_{\mathcal{G}}$, its groundstate may not be invariant $U_{\mathcal{G}}\ket{{\rm gs}(g)}\neq\ket{{\rm gs}(g)}$ as  the microscopic parameters $g$ are modified across a critical point  $g_{\rm c}$. This  leads to the so-called  symmetry-breaking phase transitions,  which can be characterized by  local order parameters. During the last decades, {\it topology} has proved to be   yet another fundamental driving force  in condensed matter, leading to partially uncharted territories with a new form of matter that cannot be described by Landau symmetry breaking, nor by local order parameters,  requiring instead the use of certain topological invariants: {\it topological matter}~\cite{review_wen_top_phases_matter}. A current challenge of modern condensed-matter physics is to understand the interplay of these three key ingredients: symmetry, topology, and many-body effects, trying to  classify  all the possible orderings and the new physical phenomena that they entail.

The theoretical prediction of topological insulators~\cite{kane_mele_z2,kane_mele_qsh,BHZ_model}, followed by  their experimental realization~\cite{qsh_exp,3D_TRI_top_ins_exp}, have caused a revolution in the subject  of topological matter. In particular, these seminal contributions showed that the combination of symmetry and topology brings in an unexpected twist in the standard  band theory of solids, establishing topological matter as one of the most active research areas of condensed matter~\cite{ti_review_1,ti_review_2}. Let us note that the symmetries in topological insulators are not related to the  previous  unitaries $U_{\mathcal{G}}$ commuting with the Hamiltonian, but correspond instead to discrete time-reversal $\mathcal{T}$ and particle-hole $\mathcal{C}$ symmetries, and the combination thereof $\mathcal{S}=\mathcal{T}\circ\mathcal{C}$, which allow to classify Hamiltonians according to ten generic symmetry classes.  In this context, topological insulators might be considered as the ``symmetric counterparts'' of the integer quantum Hall effect~\cite{vonKlitzing_1986}: they also display  topological quantization~\cite{qhe_top_invariant} and  edge/surface states~\cite{qhe_edge_states} that are robust to perturbations within a particular symmetry class~\cite{class_ti_prb,class_ti_kitaev,classification_review}.   Let us emphasize that  symmetry plays a completely different  role in this scenario. Whereas standard phases  can be understood from  the pattern of broken symmetries,   topological insulators require topological invariants that remain quantized in the presence of perturbations that respect the  particular symmetry. 

Topological insulators can be understood by upgrading the aforementioned electronic band theory  to a topological band theory, and can thus be classified as a single-particle effect. An interesting and active line of research is to explore    quantum many-body effects in topological insulators as interactions among the electrons are switched on, and inter-particle correlations build up, leading to  {\it correlated topological insulators}~\cite{review_int_top_ins,review_int_top_ins_II}. Another possible and, arguably, more interesting question is to study  if interactions allow for new topological phenomena not present in the single-particle limit.

Interactions may stabilize a topological phase that cannot be adiabatically connected to  the  free-fermion limit by gradually decreasing the interactions. This leads to the so-called  {\it interaction-induced topological insulators}. For instance, initial mean-field studies~\cite{int_QAH_honeycomb_MF} and subsequent works~\cite{Weeks_2010,castro_2011,grushin_2013,dauphin_2012} showed that the Haldane topological insulator~\cite{QAH_haldane} can be dynamically generated from a  semi-metal by including interactions. For instance, for spinless fermions in a honeycomb lattice with Hubbard-type interactions, the anti-unitary $\mathcal{T}$ symmetry can be broken   for sufficiently-strong electron-electron interactions, leading to a topological Chern insulator within its particular symmetry class~\cite{int_QAH_honeycomb_MF}. Although this idea was ruled out by subsequent numerical studies that found a density-wave-type Landau order in detriment of the Chern insulator~\cite{absence_int_QAH_honeycomb_exact_diag,absence_int_QAH_honeycomb_exact_dmrg}, its essence turned out to be  correct,  as interaction-induced Chern insulators can be found  starting from  different  non-interacting models~\cite{interaction_induced_top_insulators,QAH_kagome_dmrg,QAH_checkerborad_exact_diag,QAH_checkerborad_exact_dmrg}. Moreover,  unitary symmetries can also be spontaneously broken, such that the clear-cut division between the role of symmetry breaking in standard Landau-ordered phases and  symmetry  protection in topological insulators dissapears. This can lead to {\it symmetry-breaking topological insulators}~\cite{SBT_fermions_triang_lattice}, where the topological insulating behavior within a particular symmetry class  sets in by the spontaneous  breaking of a different unitary symmetry. In this situation, the ground-state can  simultaneously display  topological features, characterized by a topological invariant, and Landau ordering,  characterized by a local order parameter. For instance, spinless fermions in a triangular lattice with Hubbard-type interactions can lead to a phase displaying both density-wave-type order and a fractional Chern insulator~\cite{SBT_fermions_triang_lattice}.

At this point, we remark that this type of correlated topological matter is not restricted to fermionic topological insulators, but  also occurs for  spin and bosonic models, which are altogether referred to as {\it symmetry-protected topological} (SPT) phases~\cite{spt_phases_review}. In fact, the spin-1 Heisenberg chain~\cite{haldane_phase_heisenberg,aklt_haldane_edge_spins} is arguably the first instance of a correlated SPT phase without a single-particle non-interacting counterpart~\cite{haldane_spt,entanglement_spectrum}. Additionally, bosonic SPT phases are important in the abstract classification~\cite{SPT_int_bosons_general_theory,SPT_int_bosons_chern_simmons} of SPT phases of matter beyond 1D~\cite{classification_one_dimension_1,classification_one_dimension_2}. For this reason,  although the focus still lies on  correlated fermionic SPT phases, the bosonic case is gradually raising more attention from the community as  microscopic lattice models are put forth~\cite{integer_QHE_int_bosons_chern_simmons,integer_QHE_int_bosons_square_lattice_exact_diag}. From this perspective, the promising prospects of quantum simulations~\cite{quantum_simulation_1} with ultra-cold neutral atoms in optical lattices~\cite{bloch} gives a further motivation, as  lattice models of interacting bosons such as the Bose-Hubbard model (BHM) ~\cite{superfluid_mott_proposal,superfluid_mott}, can now be experimentally realized and probed to unprecedented levels of precision. Moreover, topological band-structures have already been loaded with weakly-interacting ultra-cold bosons, and the topological features have been  probed by various means~\cite{rice_mele_cold_bosons_exp,hofstadter_cold_bosons_exp,hofstadter_cold_bosons_exp_bis,hofstadter_cold_bosons_chern_exp,Tai2017}. As recently shown in~\cite{integer_QHE_int_bosons,ssh_bose_hubbard}, these topological band-structures can lead to interaction-induced SPT  phases as the boson-boson interactions are increased towards the strongly-correlated regime. It is interesting to note that other interaction-induced bosonic topological insulators have been theoretically predicted to exist already at the weakly-interacting regime~\cite{top_superfluid_square_lattice_bosons_bogoliubov_mf_weak_interactions,QAH_lieb_lattice_int_bosons_bogoliubov_mf_weak_interactions}.

The goal of this paper is to present a through description of a lattice model that can host a bosonic groundstate corresponding to an interaction-induced symmetry-breaking topological insulator. In particular, we will show that a one-dimensional bosonic SPT phase arises at finite boson-boson interactions, and cannot be adiabatically connected to the non-interacting system. This SPT phase occurs via the spontaneous breaking of lattice  translational invariance, which also produces a long-range order in the bond density of bosons. Therefore, the bosonic groundstate combines a topological-insulating behavior with Landau-type order, leading to a particular instance of symmetry-breaking topological insulators: a {\it topological bond-order wave} (TBOW). To  our knowledge, our results constitute the first instance of a bosonic interaction-induced symmetry-breaking topological insulator. 

The article is organized as follows. In Sec.~\ref{sec:model}, we introduce the $\mathbb{Z}_2$ Bose-Hubbard model and explain its connection with general lattice field theories, such as gauge theories, as well as with fermion-phonon lattice models in condensed matter. Similarly to the latter, our model exhibits a spontaneous breaking of the translational symmetry, giving rise to long-range Landau-type order. In Sec.~\ref{sec:spt_phases}, we study this phenomenon in detail, focusing first on the hardcore boson limit. Using a Born-Oppenheimer approximation for quasi-adiabatic $\mathbb{Z}_2$ fields, we predict the opening of a single-particle gap at half filling, associated to a dimerization in the structure of the $\mathbb{Z}_2$ fields. We show that one of the symmetry-broken sectors of this ordered phase  leads to a topological hardcore-boson insulator: a TBOW,  which is characterized by a quantized topological invariant, the Zak phase. We check that this TBOW phase survives in the softcore regime as  the interaction strengths are reduced, and show how the size of the gap decreases, suggesting that a quantum phase transition may occur at finite interactions that would prove that the TBOW phase  is a bosonic instance of an interaction-induced symmetry-breaking topological insulator. In Sec.~\ref{sec:dmrg}, we test these predictions numerically using the density matrix renormalization group algorithm. We give several signatures to characterize the TBOW as a SPT phase, and discuss the existence of fractional many-body edge states. We also analyze the phase transition between the topological insulator and a non-topological superfluid phase for small interactions, presenting a phase diagram of the mode, and showing the importance of strong correlations to stabilize the topological phase. Finally, in Section~\ref{sec:conclusions} we summarize our results.

\section{The $\mathbb{Z}_2$ Bose-Hubbard model}
\label{sec:model}

In this section, we describe the bosonic lattice model under study within the more general context of lattice field theories~\cite{lft_book}. In lattice field theories, one is typically concerned with the discretization of a  local quantum field theory with a matter sector that consists of particles whose interactions are carried by the excitations of some gauge field. This discretization proceeds by placing the matter fields, typically described by relativistic fermions, on the sites of a lattice that serves as a scaffolding of space (or space-time), while the carriers, typically described by  gauge bosons, reside on the lattice links. Let us emphasize, however, that other interesting situations have also been explored in the literature.

For instance, one can place $\mathbb{Z}_2$ fields in the links, and study the so-called Ising lattice gauge theories~\cite{pure_z2_gauge, z2lgt}, which may interact with a bosonic matter sector residing on the lattice sites~\cite{z2_gauge_bosons, bosonic_lgt}. More recently,  gauge theories  of (2+1) relativistic fermions coupled to $\mathbb{Z}_2$ fields have  provided a very rich playground to test the phenomenon of confinement-deconfinement transitions in a condensed-matter scenario~\cite{z2_gauge_fermions,z2_gauge_fermions_no_plaquette, z2_gauge_fermions_hubbard}. Let us note that one can be  more general, and study fermionic lattice field theories under $\mathbb{Z}_2$ fields where the gauge symmetry is explicitly broken. This can also lead to exotic strongly-correlated behavior~\cite{z2_field_fermions_no_gauge,z2_field_fermions_no_gauge_dyn_mass_generation} and, in the context of the present manuscript,  to dynamical generation of topological masses that can stabilize  an  interaction-induced SPT phase~\cite{z2_field_fermions_no_gauge_dyn_mass_generation_spt_phase}. 

\begin{figure}[t]
  \centering
  \includegraphics[width=1.0\linewidth]{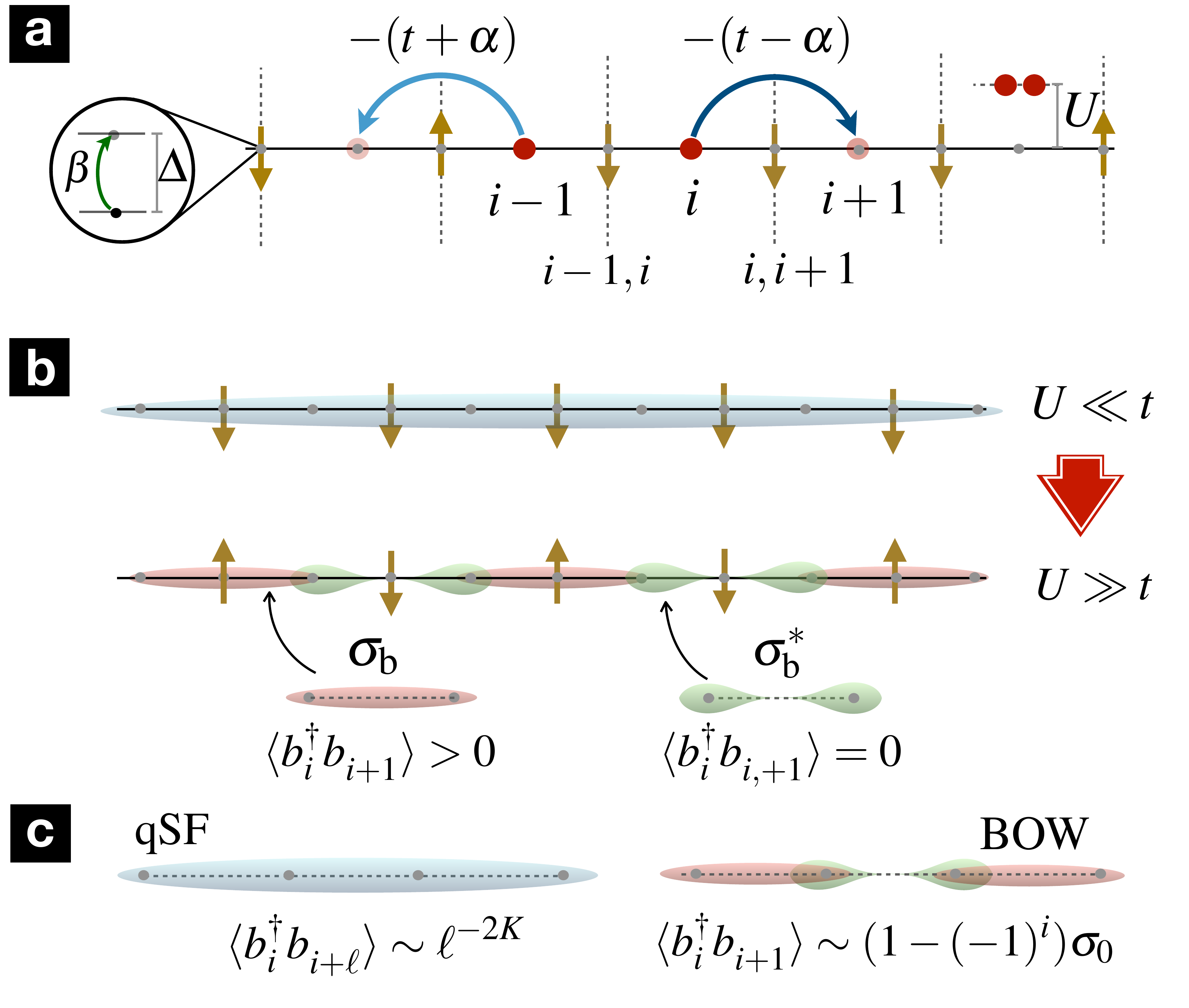}
\caption{\label{fig:scheme} {\bf Scheme of the bosonic Peierls transition:} {\bf (a)} Representation of the lattice model of bosons coupled to an Ising field according to Eqs.~\eqref{eq:sb_hamiltonian}-\eqref{eq:spin_h}. Bosons (red dots) reside on the lattice sites, where they interact with strength $U$, and can tunnel through the links with a $\mathbb{Z}_2$-valued tunneling strength $-t\pm\alpha$, which depends on the configuration of the Ising field on the links (yellow arrows). These fields can be represented by a two-level system (inset) with energy difference $\Delta$, and spin-flipping strength $\beta$.  {\bf (b,c)} Interaction-induced Peierls transition between a bosonic quasi-superfluid (qSF)  in a homogeneous Ising backround, and a bond-ordered wave (BOW) in a N\'eel-orfdered type background, such that  translational invariance is spontaneously broken. The BOW phase can be understood as an alternation of bonding $\sigma_{\rm b}$  and anti-bonding $\sigma_{\rm b}^*$  units with a different distribution of the bosons within the links/bonds. The homogeneous qSF phase is characterized by an algebraic decay of correlations.}
\end{figure}

In this article, we study a (1+1) Hamiltonian lattice field theory of bosons coupled to  $\mathbb{Z}_2$ fields, and introduce terms that explicitly break the gauge symmetry. As we will  discuss in detail below, these ingredients  provide a  rich playground to explore the aforementioned TBOW. In this reduced dimensionality~\cite{z2_gauge_fermions_1d}, the $\mathbb{Z}_2$ gauge theory of bosons~\cite{z2_gauge_bosons} becomes
\begin{equation}
\label{eq:sb_hamiltonian}
H_{\mathsf{Z}_2}=-\alpha\sum_i\left({b}^\dagger_i{\sigma}^{z\vphantom{\dagger}}_{i,i+1}{b}^{\vphantom{\dagger}}_{i+1}+\text{H.c.}\right)+\beta\sum_i{\sigma}^x_{i,i+1},
\end{equation}
where we have introduced the discretized bosonic operators $b_{i}^{\vphantom{\dagger}},b_i^\dagger$ on the sites of a 1D crystal, and the $\mathbb{Z}_2$ Ising field is described by Pauli matrices ${\sigma}^{z\vphantom{\dagger}}_{i,i+1},{\sigma}^{x\vphantom{\dagger}}_{i,i+1}$ that reside on the links/bonds of the chain. In this Hamiltonian, $\alpha>0$ represents the strength of a gauge-invariant tunneling of bosons, and $\beta>0$ stands for a transverse field that induces a spin-flip dynamics on the $\mathbb{Z}_2$ fields (see Fig.~\ref{fig:scheme}{\bf (a)}). This Hamiltonian~\eqref{eq:sb_hamiltonian} constitutes the  simplest bosonic Ising lattice  gauge theory, as it displays a local $\mathbb{Z}_2$ symmetry:  the bosons transform as $b_i\to{\rm e}^{{\rm i}\varphi_i}b_i$,  $b_i^\dagger\to{\rm e}^{-{\rm i}\varphi_i}b_i^\dagger$, while the Ising fields fulfill  ${\sigma}^{x\vphantom{\dagger}}_{i,i+1}\to\sigma^{x\vphantom{\dagger}}_{i,i+1}$,  ${\sigma}^{z\vphantom{\dagger}}_{i,i+1}\to{\rm e}^{{\rm i}\varphi_i}{\sigma}^{z\vphantom{\dagger}}_{i,i+1}{\rm e}^{-{\rm i}\varphi_{i+1}}$, where $\varphi_i\in\{0,\pi\}$ is a local $\mathbb{Z}_2$-valued phase.

As advanced below, interesting scenarios for the study of correlated topological matter can arise by considering additional terms that explicitly break the gauge symmetry. In our context, we consider a paradigmatic model of strongly-correlated bosons, the so-called Bose-Hubbard model~\cite{bose-hubbard_1,bose-hubbard_2}
\begin{equation}
\label{eq:bose-hubbard}
H_{\mathsf{BH}}=-t\sum_i \left({b}^\dagger_i{b}^{\vphantom{\dagger}}_{i+1}+\text{H.c.}\right)+\frac{U}{2}\sum_i {n}_i({n}_i-1),
\end{equation}
where ${n}_i={b}^\dagger_i{b}^{\vphantom{\dagger}}_i$. Therefore, in addition to the $\mathbb{Z}_2$-dressed tunneling of Eq.~\eqref{eq:sb_hamiltonian}, the bosons also have a bare tunneling of strength $t$, and contact interactions of strength $U$. Finally, we  introduce the simplest-possible gauge-breaking  term that modifies the spin dynamics
\begin{equation}
\label{eq:spin_h}
H_{\Delta}=\frac{\Delta}{2}\sum_i{\sigma}^z_{i,i+1},
\end{equation}
 where $\Delta$ stands for the energy difference between the local configurations $\ket{{\uparrow}_{i,i+1}},\ket{{\downarrow}_{i,i+1}}$  of the $\mathbb{Z}_2$ field.
 
 Altogether, the Hamiltonian containing the  terms~\eqref{eq:sb_hamiltonian}-\eqref{eq:spin_h}
 \beq
 \label{eq:z2_bhm}
 H_{\mathsf{Z_2BH}}=H_{\mathsf{Z}_2}+H_{\mathsf{BH}}+H_{\Delta}
 \eeq
   forms the model of strongly-correlated lattice bosons under $\mathbb{Z}_2$ fields ($\mathbb{Z}_2$BHM) that is the subject of our work.
   
   Recently, we have studied this model in the context of {\it bosonic Peierls transitions}~\cite{paper}, as the $\mathbb{Z}_2$ field can be considered as  a simplified version of a dynamical lattice with the vibrational  phonons  substituted by discrete Ising variables. From this perspective, we have shown that bosonic analogues of some of  the  rich phenomenology of density waves in strongly-correlated materials~\cite{dw_1d_chains_review}, customarily described in terms of fermion-phonon models, can also be found in the $\mathbb{Z}_2$BHM. These transitions are characterized by the spontaneous breaking of the translational invariance of the $\mathbb{Z}_2$ field, and the development of a bond-ordered wave for the bosons that displays a periodic modulation of the density of bosons along the links of the chain (see Fig.~\ref{fig:scheme}{\bf (b)}-{\bf (c)}).  This phenomena is reminiscent of the standard Peierls transition in   1D metals~\cite{peierls}, where the fermionic statistics and the presence of a Fermi sea are sufficient to drive this effect (i.e. a gap opening at the Fermi surface provides the required energy  to distort the lattice). Depending on the fermionic occupation, different distortion patterns can emerge. The associated order is characterized by a wavelength $\lambda_0$, which is inverse of the density $\rho$, $\lambda_0 = 1/\rho$. Although these notions are absent in our bosonic model, we have found that similar effects can still appear for sufficiently strong Hubbard  interactions (see Fig.~\ref{fig:scheme}{\bf (b)}-{\bf (c)}).

The focus of our previous work~\cite{paper} was  the elucidation of this bosonic Peierls mechanism for different bosonic densities, as well as the study of topological solitons (i.e. kinks) interpolating between different bond-density modulations at commensurate fillings. The latter are a direct consequence of the degeneracies associated to the symmetry-breaking process.  In the present work, we focus on a different topological aspect: we give compelling evidence that, at half-filling, the  bond-ordered wave (BOW) caused by the bosonic Peierls mechanism corresponds to an interaction-induced SPT phase. This topological phase occurs simultaneously with the Landau symmetry-breaking order described by the bond-density modulation. We will show that, in addition to the BOW phases of Figs.~\ref{fig:scheme}{\bf (b)}-{\bf (c)}), the pattern of broken symmetry also allows for topological bond-ordered waves (TBOW) phases that display all the characteristics of an interaction-induced SPT phase: ($\mathsf{1}$)  the appearance of non-vanishing bulk topological invariants for the many-body interacting model, and ($\mathsf{2}$) the presence of  non-trivial many-body edge states.  We will emphasize how interactions and symmetry breaking are  fundamental necessary ingredients  for  these topological effects to take place. 

\section{Topological bond-order wave: an adiabatic variational ansatz}
\label{sec:spt_phases}

In this section, we provide a thorough description of our findings supporting the existence of bosonic TBOW in our lattice model~\eqref{eq:sb_hamiltonian}-\eqref{eq:spin_h}. In order to guide the presentation  of our results, we start by discussing symmetry-protected topological phases in dimerized lattice models. We  then show how similar phases arise spontaneously in our model introduced in Eq.~\eqref{eq:z2_bhm} at half filling, as a consequence of the breaking of the translational symmetry (bosonic Peierls transition), giving rise to a topological bond-order wave (TBOW). We analize first the case of hardcore bosons, where clear analogies with the standard Peierls transition in fermion-phonon systems can be drawn, and discuss the interplay between symmetry breaking and symmetry protection. We then show how the TBOW phase extends to finite Hubbard interactions. Based on different analytical approximations, we provide arguments for the existence of a phase transition to a non-topological phase for small enough interactions, supporting our claim that strong correlations are necessary to induced the symmetry-broken topological phase.

\subsection{Symmetry-protected topological phases}

In 1D, there are two types of fermionic SPT phases with chiral/sublattice $\mathcal{S}$ symmetry, the so-called $\mathsf{BDI}$ and $\mathsf{AIII}$ topological insulators~\cite{class_ti_prb,class_ti_kitaev,chiu_2016}.  The   $\mathsf{BDI}$ class also fulfils time-reversal $\mathcal{T}$, particle-hole $\mathcal{C}$, and has a simple representative: a fermionic tight-binding model with dimerized  tunnelings
\beq
\label{eq:huckel}
H_{\mathsf{BDI}}(t,\delta )=\sum_i\left(-t(1+(-1)^i\delta )c_i^\dagger c_{i+1}^{\vphantom{\dagger}}+{\rm H.c.}\right),
\eeq
where $c_i^\dagger, c_i^{\vphantom{\dagger}}$ are fermionic creation-annihilation operators, and the tunneling strengths are distributed according to a dimerized pattern $\{t(1-\delta ),t(1+\delta ),\cdots,t(1+\delta ), t(1-\delta )\}$. This model is known as the H\"{u}ckel model~\cite{huckel_model} in  the context of organic chemistry and conjugated polymers~\cite{barford}. The model is periodic with a two-atom unit cell $(A,B)$ and, in momentum space, the Hamiltonian reads

\beq
H_{\mathsf{BDI}}(t,\delta )=\sum_k C^\dagger_k \; \mathbf{n}(k)\cdot \boldsymbol{\sigma} \; C_k,
\eeq
where we have introduced $C_k=(c_A(k),c_B(k))^{\rm t}$ and  $\mathbf{n}(k)=-t(1-\delta+(1+\delta)\cos k,(1+\delta)\sin k,0)$ and $\boldsymbol{\sigma}=(\sigma_x,\sigma_y,\sigma_z)$ is the Pauli vector. At half filling, the groundstate  corresponds to a $\mathsf{BDI}$ topological insulator in the interval $\delta \in(0,2)$, and to a trivial band insulator elsewhere. These two phases can be distinguished by a global topological invariant, the winding number $\nu$~\cite{chiu_2016,Asboth_2016}, defined as

\beq
\label{eq:winding}
\nu=\frac{1}{2\pi}\int_{-\pi}^{\pi}dk \left(\mathbf{n}(k)\times \partial_k \mathbf{n}(k)\right)_z.
\eeq

The winding number is a $\mathbb{Z}$ topological invariant and, in the present case, has value $1$ in the interval $\delta \in (0,2)$ (i.e. in the $\mathsf{BDI}$ topological phase) the  and $0$ elsewhere. Let us emphasize that, in general, this quantity is not limited to $0$ or $1$: it is in fact possible to have greater winding numbers by considering Hamiltonians with long-range hopping that preserve the chiral symmetry~\cite{maffei2018}. For hard-wall boundary conditions, the $\mathsf{BDI}$ topological insulator exhibits edge states protected by the topology of the system, this is, they are robust against perturbations that respect the chiral symmetry and do not close the gap. There exists a relation between the number of edge states and the winding number, called the bulk-edge correspondence~\cite{chiu_2016,Asboth_2016}: at each edge of the system, the number of edge states is equal to the winding number.  The  H\"{u}ckel model has been generalized to the spinful Fermi-Hubbard H\"{u}ckel model, where the bulk-edge correspondence has been studied by means of the entanglement spectrum and the presence of topological edge states~\cite{Wang_2015, Ye_2016,Barbiero_2018}.

Interestingly, chiral systems have another topological invariant, the Zak phase~\cite{zak_1989} 
\beq
\label{eq:zak_phase}
\varphi_\text{Zak}=\int_{\rm BZ}{\rm d}k\,\mathcal{A}_-(k),
\eeq
written in terms of  the Berry connection $\mathcal{A}_-(k)=\bra{\epsilon_{k,-}}\ii\partial_k\ket{\epsilon_{k,-}}$ over the occupied band $\ket{\epsilon_{k,-}}$. The Zak phase is a $\mathbb{Z}_2$ topological invariant and can take values $0$ or $\pi$, modulo $2\pi$. The Zak phase can be directly related to the polarization~\cite{resta_1994,Asboth_2016}, a physical observable, which basically measures the center of mass per unit cell. The two invariants, the Zak phase and the winding number, can be related through $\varphi_\text{Zak}=\pi \, [\nu (\text{mod} \, 2)]$~\cite{chiu_2016,Asboth_2016}. Therefore, in the case of  the H\"{u}ckel model~\eqref{eq:huckel}, the Zak phase and the winding number coincide. In the context of topological insulators~\cite{Asboth_2016}, Eq.~\eqref{eq:huckel} is typically referred to as the Su-Schrieffer-Hegger (SSH) Hamiltonian~\cite{ssh_model}. However, we note that the original SSH model is a fermion-boson model describing a metal coupled to the vibrational modes of a dynamical lattice (i.e. acoustic phonons). In contrast to the dimerized H\"{u}ckel model~\eqref{eq:huckel}, the SSH Hamiltonian yields a truly quantum many-body problem, and an archetype of strongly-correlated behavior~\cite{dw_1d_chains_review} that can display a  variety of exotic effects~\cite{ssh_polymers} not present in Eq.~\eqref{eq:huckel}.

Although the H\"{u}ckel model was initially introduced for fermions, this model can also be studied for a Bose-Hubbard model with dimerized tunnelings
\beq
\label{eq:dBHM}
H_{\mathsf{dBH}}=\sum_i\!\left(-t(1+(-1)^i\delta )b_i^\dagger b_{i+1}^{\vphantom{\dagger}}+{\rm H.c.}\!\right)+\frac{U}{2}\sum_in_i(n_i-1).
\eeq
In the hardcore-boson  limit $U\to\infty$, one can apply the Jordan Wigner transformation~\cite{Jordan1928}, $b_i\to\ee^{\ii\pi\sum_{j<i}c_j^\dagger c_j^{\phantom{\dagger}}}c_i^{\phantom{\dagger}}$, $b_i^\dagger b_i^{\phantom{\dagger}}\to c_i^\dagger c_i^{\phantom{\dagger}}$ and recover Eq.~\eqref{eq:huckel}. The regime of finite interactions is also very interesting from the topological viewpoint, as discussed in Ref.~\cite{ssh_bose_hubbard}. When passing from the hardcore boson limit to finite interactions, the system looses the chiral symmetry. Nevertheless, the Zak phase remains quantized due to the presence of the inversion symmetry. This model therefore passes from a $\mathsf{BDI}$ topological insulator with a $\mathbb{Z}$ invariant to an inversion-symmetric SPT phase with a $\mathbb{Z}_2$ topological invariant. It is also important to notice that in this case, the bulk-edge correspondence can be violated. 

In the context of the present paper, we are interested in the bosonic analogue of symmetry-breaking bond-order-wave formation in polymers~\cite{ssh_polymers}, and its interplay with bosonic SPT phases. This phenomena require a  dynamical lattice, and cannot thus be accounted for via the dimerized Bose-Hubbard model of Eq.~\eqref{eq:dBHM}. The goal of this section is to show that our $\mathbb{Z}_2$BHM of strongly-correlated bosons coupled to $\mathbb{Z}_2$ fields~\eqref{eq:z2_bhm}, where the role of the dynamical lattice is played by the discrete Ising spins, can indeed provide such a scenario.

\subsection{Symmetry-breaking topological phases}

\subsubsection{Hardcore bosons coupled to $\mathbb{Z}_2$ fields}
\paragraph{Born-Oppenheimer groundstate ansatz.--}

Let us now elaborate on the aforementioned analogy of the groundstate behavior of the $\mathbb{Z}_2$BHM~\eqref{eq:z2_bhm} to the standard Peierls transition and  SSH-type phenomena. As  discussed above for the simpler dimerized model~\eqref{eq:dBHM}, the hardcore-boson limit $U\to\infty$ is a good starting point to draw these analogies, since the strongly-interacting bosons of Eq.~\eqref{eq:z2_bhm} can be transformed into free spinless fermions coupled to the $\mathbb{Z}_2$ fields. By applying the previously-introduced Jordan-Wigner transformation~\cite{Jordan1928} in the hardcore-boson limit, we find 
\begin{equation}
\label{eq:z2bhm_hamiltonian_fermions}
H_{\mathsf{Z}_2\mathsf{BH}}^{U\to\infty}=\!\!\sum_i\!\!\left(\!\!-(t+\alpha{\sigma}^{z\vphantom{\dagger}}_{i,i+1}){c}^\dagger_i{c}^{\vphantom{\dagger}}_{i+1}+\frac{\beta}{2}{\sigma}^x_{i,i+1}+\frac{\Delta}{4}{\sigma}^z_{i,i+1}+\text{H.c.}\!\!\right)\!\!.
\end{equation}
In this context, one may readily notice  that  a backround  N\'eel-type anti-ferromagnetic ordering of the  $\mathbb{Z}_2$ fields $\ket{\cdots ,{\uparrow}_{i-1,i},{\downarrow}_{i,i+1}, {\uparrow}_{i+1,i+2},\cdots}$ introduces   a dimerized pattern of the fermionic tunneling strengths (see Fig.~\ref{fig:scheme}{\bf (a)}). This would constitute a $\mathbb{Z}_2$ analogue of the dimerized lattice distortion that underlies  the  fermionic Peierls instability at half filling~\cite{peierls}. However, the dynamics of the $\mathbb{Z}_2$ fields differs  from the  acoustic vibrational branch of the original   (SSH) Hamiltonian~\cite{ssh_model}, which can lead to crucial differences. 

In order to understand these differences, we shall focus on the quasi-adiabatic regime  $\beta\ll t$, where   the $\mathbb{Z}_2$ fields are much slower than  the lattice bosons. Following a Born-Oppenheimer-type reasoning,  we consider that the hardcore bosons  adapt instantaneously to the background static spins. In this way, they provide an effective potential energy for the $\mathbb{Z}_2$ fields which is used, in turn,  to determine the groundstate spin configuration. In our context, this can be formalized by means of the following variational ansatz
\beq
\label{eq:BO_var_ansatz}
\ket{\Psi_{\rm gs}(\{d_{\boldsymbol{n}},\bs\theta\})}=\ket{\psi_{\rm f}(\{d_{\boldsymbol{n}}\})}\bigotimes\ee^{-\ii\sum_i\frac{\theta_{i,i+1}}{2}\sigma_{i,i+1}^y}\ket{-}_{\rm s}
\eeq
where $\ket{\psi_{\rm f}(\{d_{\boldsymbol{n}}\}}=\sum_{\boldsymbol{n}}d_{\boldsymbol{n}}\ket{\boldsymbol{n}}_{\rm f}$ is a generic fermionic wavefunction. This wavefunction is defined by  the set of variational amplitudes $\{d_{\boldsymbol{n}}\}$ in the Fock basis $\ket{\boldsymbol{n}}_{\rm f}=\ket{n_1,\cdots,n_N}_{\rm f}$ with $n_i\in\{0,1\}$ fermions at site $ i\in\{1,\cdots,L\}$. On the other hand, this ansatz~\eqref{eq:BO_var_ansatz} describes the slow $\mathbb{Z}_2$ fields  in terms of spin coherent states with variational angles $\bs \theta=(\theta_{1,2}...\theta_{i,i+1}...)$, and reference state $\ket{-}_{\rm s}=\otimes_i(\ket{\uparrow_{i,i+1}}-\ket{\downarrow_{i,i+1}})/\sqrt{2}$.

Our Born-Oppenheimer-type variational ansatz \eqref{eq:BO_var_ansatz} can be applied at arbitrary boson filling, where  complex $\mathbb{Z}_2$ fields patterns ({\it i.e.} solitonic, incommensurate) may arise due to Peierls instability~\cite{paper}. Here we focus on the half-filled case in which, according to the previous discussion, a Peierls instability can lead to the doubling of the unit cell. Therefore, for periodic boundary conditions, it suffices to consider only two variational angles, namely $\bs\theta=(\theta_A,\theta_B)$ for the links joining odd-even (even-odd) lattice sites. As detailed in Appendix~\ref{sec:app:details}, the variational problem reduces to the minimization of the following ground-state energy
\beq
\begin{split}
\label{eq:var_energy}
\epsilon_{\rm gs}(\bs\theta)=&-\frac{2}{\pi}
t(\bs\theta)
\mathsf{E}\big(1-\delta^2\!(\bs\theta)\big)\\
&+\frac{\Delta}{4}(\sin\theta_A+\sin\theta_B)-\frac{\beta}{2}(\cos\theta_A+\cos\theta_B),
\end{split}
\eeq
where we have introduced $t(\bs\theta)=t+\frac{\alpha}{2}\big(\sin\theta_A+\sin\theta_B\big)$, $\delta(\bs\theta)=\alpha(\sin\theta_A-\sin\theta_B)/(2t+\alpha(\sin\theta_A+\sin\theta_B))$, and  $ \mathsf{E}(x)=\int_0^{\pi/2}{\rm d}k(1-x\sin^2k)^{1/2}$ is the complete elliptic integral of the second kind. Note that this minimization  shall yield the particular angles $\bs \theta^\star=(\theta^\star_A,\theta^\star_B)$, which  determine the $\mathbb{Z}_2$-field background experienced by the  hardcore bosons.

\begin{figure*}[t]
  \centering
  \includegraphics[width=0.9\linewidth]{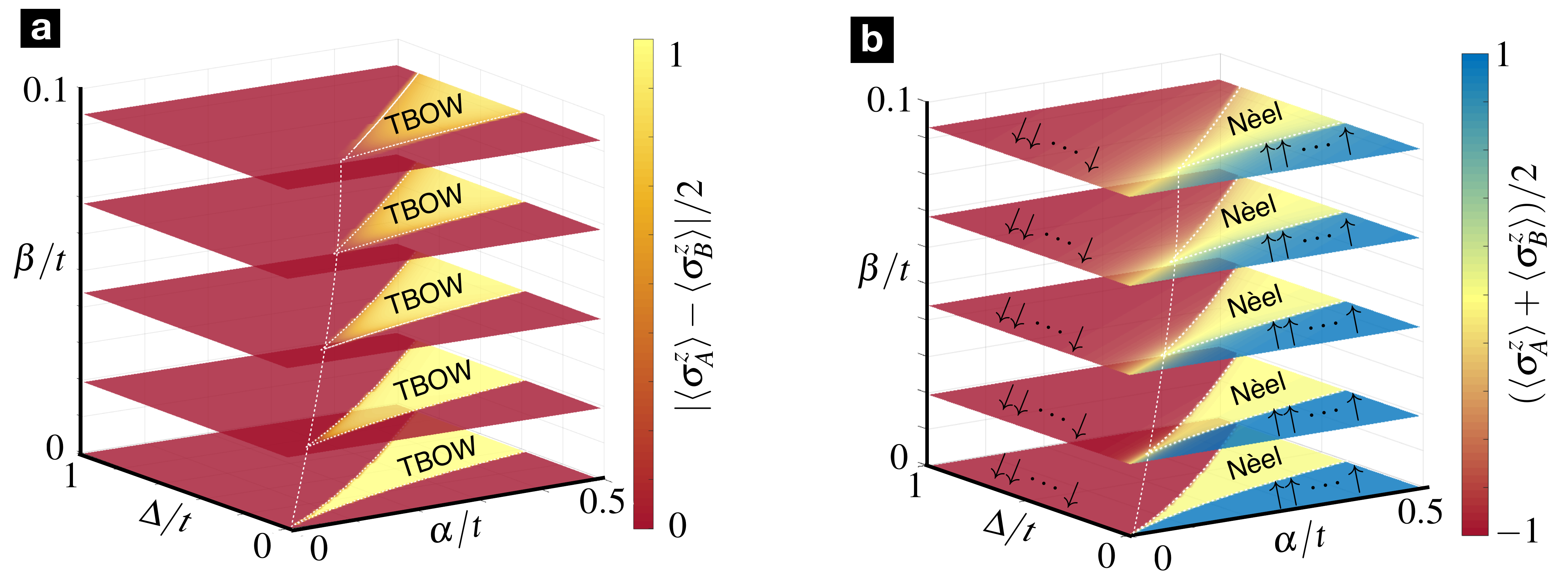}
\caption{\label{fig:BO_phase_diagram} {\bf Born-Oppenheimer phase diagram of the $\mathbb{Z}_2$ Bose-Hubbard model:} {\bf (a)} Representation of the magnetization difference $|\langle\sigma^z_A\rangle-\langle\sigma^z_B\rangle|$ between the even-odd sub-lattices for the variational groundstate obtained by minimizing Eq.~\eqref{eq:var_energy} for different paramters $(\Delta/t,\alpha/t,\beta/t)$. The yellow region corresponds to the two possible N\'eel configurations of the $\mathbb{Z}_2$ fields leading to a bond-density order parameter, and a symmetry-broken TBOW. {\bf (b)} Representation of the magnetization sum $|\langle\sigma^z_A\rangle+\langle\sigma^z_B\rangle|$ between the even-odd sub-lattices for the variational groundstate obtained by minimizing Eq.~\eqref{eq:var_energy} for different paramters $(\Delta/t,\alpha/t,\beta/t)$. The red and blue regions correspond to the fully polarized  configurations of the $\mathbb{Z}_2$ fields, which do not induce any modulation of the bosonic tunnelings, and thus lead to trivial insulators. }
\end{figure*}

At this variational level, we can draw a clear  analogy between  the $\mathbb{Z}_2$BHM~\eqref{eq:z2_bhm}  and the standard Peierls instability of the SSH model~\cite{ssh_model}.  In the SSH model,  the energy reduction of the  fermions due to a gap opening in the 1D metal compensates for the  elastic energy increase of the lattice distortion (i.e. static limit of the acoustic branch)~\cite{peierls}. In our case, the quadratic elastic energy of the standard Peierls problem is substituted by a  trigonometric function describing the energy of the $\mathbb{Z}_2$-field background (see the second line of Eq.~\eqref{eq:var_energy}). A direct consequence of this difference is that, whereas the 1D metal of the SSH model is always unstable towards a BOW phase at $T=0$, the $\mathbb{Z}_2$ Peierls instability of our hardcore bosons  does indeed depend on the ratio $\Delta/\alpha$, such that one can observe a Peierls transition even at zero temperatures (i.e. Peierls quantum phase transition~\cite{paper}).

 \paragraph{Born-Oppenheimer excitation ansatz.--} In order to carry further this analogy, we discuss the gap opening in the $\mathbb{Z}_2$BHM~\eqref{eq:z2_bhm}, which requires  generalizing the Born-Oppenheimer ansatz~\eqref{eq:BO_var_ansatz} to account for the low-energy single-particle excitations. In a  first step, we consider that the spin fluctuations about the $\mathbb{Z}_2$-field background are small, and introduce a spin-wave-type~\cite{spin_wave} formulation based on a Holstein-Primakoff transformation~\cite{holstein_primakov}, namely
\beq
\label{eq:spin-wave}
\begin{split}
\sigma_{i,i+1}^z&\approx\cos\theta_{i,i+1}^\star\left(a^{\phantom{\dagger}}_i+a^{{\dagger}}_i\right)-\sin\theta_{i,i+1}^\star\left(2a^{{\dagger}}_ia^{\phantom{\dagger}}_i-1\right),\\
\sigma_{i,i+1}^x&\approx\sin\theta_{i,i+1}^\star\left(a^{\phantom{\dagger}}_i+a^{{\dagger}}_i\right)+\cos\theta_{i,i+1}^\star\left(2a^{{\dagger}}_ia^{\phantom{\dagger}}_i-1\right),
\end{split}
\eeq 
where $a^{{\dagger}}_i,a^{\phantom{\dagger}}_i$ are bosonic creation-annihiliation operators for the  excitations of the $\mathbb{Z}_2$  fields localized at link $(i,i+1)$.

In a second step,  we introduce a  family of single-particle excitations over the previous variational ground-state $\ket{\Psi_{\rm gs}(\{{d}_{\boldsymbol{n}},\bs\theta^\star\})}$ obtained from Eq.~\eqref{eq:BO_var_ansatz} by setting to the optimum variational angles $\bs \theta^\star=(\theta^\star_A,\theta^\star_B)$, namely
\beq
\label{eq:BO_var_ansatz_exc}
\ket{\Psi_{\rm exc}(\bs\eta)}=\bigg(\sum_{k\in{\rm BZ}}\eta_{{\rm f},k}^{\phantom{\dagger}}\gamma_{k,+}^\dagger+\sum_{i=1}^N\eta_{{\rm s},i}^{\phantom{\dagger}} a_i^\dagger\bigg)\ket{\Psi_{\rm gs}(\{{d}_{\boldsymbol{n}},\bs\theta^\star\})},
\eeq
where  $\bs \eta=(\eta_{{\rm f},k},\eta_{{\rm s},i})$ are the variational amplitudes,  and  $\gamma^\dagger_{k,+}$ are Bogoliubov-type fermion creation operators in the single-particle  conduction band of the hardcore boson sector (see Appendix \ref{sec:app:details} for details). In this case, the variational functional for the excitation energies depend on
\beq
\label{eq:qpart_energies}
\begin{split}
\epsilon^{\rm f}_{k}(\bs\theta^\star)&=2t(\bs\theta^\star)\sqrt{\cos^2q+\delta^2(\bs\theta^\star)\sin^2q},\\ 
\epsilon^{\rm s}_{2i-1}(\bs\theta^\star)&=2\beta\cos\theta_A^\star-\sin\theta_A^\star\left(\frac{\Delta}{2}-2\alpha \mathsf{B}_A(\bs\theta^\star)\right),\\
\epsilon^{\rm s}_{2i}(\bs\theta^\star)&=2\beta\cos\theta_B^\star-\sin\theta_B^\star\left(\frac{\Delta}{2}-2\alpha \mathsf{B}_B(\bs\theta^\star)\right),\\
\end{split}
\eeq
which themselves depend on the properties of the variational groundstate, such as the  the fermionic  bond densities between odd-even sites $\mathsf{B}_A(\bs\theta^\star)=\mathsf{B}_{2i-1,2i}$,  and between even-odd sites $\mathsf{B}_B(\bs\theta^\star)=\mathsf{B}_{2i,2i+1}$, where
$
\mathsf{B}_{i,j}=\langle c_{i}^\dagger c^{\phantom{\dagger}}_{j}\rangle_{\rm gs}+{\rm c.c.}.
$

The variational minimization then yields two types of low-energy excitations: {\it (i)} delocalized fermion-like excitations with $\epsilon_{\rm exc}(\bs \theta^\star)=\epsilon^{\rm f}_{k}(\bs \theta^\star)$ $\forall k\in[-\frac{\pi}{2},\frac{\pi}{2})$, or {\it (ii)} localized spin-wave-type excitations with $\epsilon_{\rm exc}(\bs \theta^\star)=\epsilon^{\rm s}_{i}(\bs \theta^\star)$ $\forall i\in\{1,\cdots N\} $. Therefore, in our context, the gap opening is caused by a N\'eel-type alternation of the spins   $ \theta^\star_A-\theta^\star_B\neq 0$, which  leads to  $\delta(\bs \theta^\star)\neq 0$ and, according to Eq.~\eqref{eq:qpart_energies}, to the aforementioned gap opening
\beq
\label{eq:gap_opening}
\Delta\epsilon={\rm min}_k\epsilon_{\rm exc}(\bs \theta^\star)= 2t(\bs\theta^\star)|\delta(\bs \theta^\star)|>0.
\eeq

\paragraph{Adiabatic regime: Peierls transition and SPT phases.--} After introducing this variational machinery, we can explore the rich physics of the $\mathbb{Z}_2$BHM~\eqref{eq:z2_bhm} by focusing first on the adiabatic regime $\beta =0$, where various results can be obtained analytically. 
In this limit, where the spins are static, it is possible to solve analytically the variational minimization of Eq.~\eqref{eq:var_energy} for the ground-state ansatz~\eqref{eq:BO_var_ansatz}, finding two critical lines 
\beq
\label{eq:static_critical_lines}
\Delta_{\rm c}^{\pm}=\frac{4t}{\pi}\left(\tilde{\delta}\pm\mathsf{E}\left(1-\tilde{\delta}^2\right)\mp1\right),
\eeq
where we have defined  $\tilde{\delta}=\alpha/t$. These critical lines, represented in the lowest  planar sections of Fig.~\ref{fig:BO_phase_diagram}, are in perfect agreement with our previous results~\cite{paper}. For $\Delta>\Delta_{\rm c}^{+}$ ($\Delta<\Delta_{\rm c}^{-}$), the $\mathbb{Z}_2$-field background $\theta_A^\star=\theta_B^\star=\frac{\pi}{2}$ ($\theta_A^\star=\theta_B^\star=-\frac{\pi}{2}$) yields a polarized state  $\ket{{\downarrow\downarrow\cdots\downarrow}}$ ($\ket{{\uparrow\uparrow\cdots\uparrow}}$), such that the translational invariance remains intact. Instead, for $\Delta\in(\Delta_{\rm c}^{-},\Delta_{\rm c}^{+})$,  the variational minimization leads to the spontaneous breaking of the translational symmetry, yielding two possible perfectly-ordered N\'eel states, either $\ket{{\downarrow\uparrow\downarrow\uparrow\cdots\downarrow\uparrow\downarrow}}$ for $\theta_A^\star=-\theta_B^\star=-\frac{\pi}{2}$, or $\ket{{\uparrow\downarrow\uparrow\downarrow\cdots\uparrow\downarrow\uparrow}}$ for $\theta_A^\star=-\theta_B^\star=+\frac{\pi}{2}$. Let us now use the variational ansatz for the excitations~\eqref{eq:BO_var_ansatz_exc} to show that this phase transition is marked by a gap opening, as occurs for the standard Peierls instability in 1D metals.

According to our previous discussion~\eqref{eq:gap_opening},  as a consequence of $|\delta(\pm\pi/2,\mp\pi/2)|=\tilde{\delta}>0$,   a gap of magnitude $\Delta\epsilon=2t\tilde{\delta}$ will be opened. This  signals a Peierls transition accompanied by a BOW density  modulation 
\beq
\label{eq:BOW_parameter}
\begin{split}
\mathsf{B}_A\left(\!-\frac{\pi}{2},+\frac{\pi}{2}\right)=\frac{2}{\pi(1-\tilde{\delta})}\left(
\mathsf{E}(1-\tilde{\delta}^{2})-\tilde{\delta}\mathsf{K}(1-\tilde{\delta}^{2})\right),\\
\mathsf{B}_B\left(\!-\frac{\pi}{2},+\frac{\pi}{2}\right)=\frac{2}{\pi(1+\tilde{\delta})}\left(
\mathsf{E}(1-\tilde{\delta}^{2})+\tilde{\delta}\mathsf{K}(1-\tilde{\delta}^{2})
\right),
\end{split}
\eeq
where we assume the symmetry-broken state $\ket{{\downarrow\uparrow\cdots\uparrow\downarrow}}$ (for $\ket{{\uparrow\downarrow\cdots\downarrow\uparrow}}$, the expressions for the $A$ and $B$ sublattices must be interchanged), and make use of the complete elliptic integral of the first kind $\mathsf{K}(x)=\int_0^{\pi/2}{\rm d}k(1-x\sin^2k)^{-1/2}$. Note that $ \mathsf{B}_A=\mathsf{B}_B=2/\pi$ for $\tilde{\delta}\to0$, whereas in the limit $\tilde{\delta}\to1$, we recover the alternation $\mathsf{B}_A=0,\mathsf{B}_B=1$  between perfect antibonding-bonding   links   (see Fig.~\ref{fig:scheme}{\bf (b)}). For  spin-boson couplings $\alpha<t$, there will be a period-two modulation with a smaller antibonding-bonding character. Let us emphasize again that, contrary to  the SSH groundstate that is always unstable towards the Peierls insulator for arbitrary fermion-phonon couplings, the $\mathbb{Z}_2$BHM~\eqref{eq:z2_bhm} does support a Peierls transition as the spin-boson coupling is modified.

 Let us now discuss the interplay of symmetry-breaking  order  and symmetry-protected topological features in this BOW phase. The direct-product structure of our Born-Oppenheimer ansatz~\eqref{eq:BO_var_ansatz} allows us to extract an effective Hamiltonian for the hardcore boson sector   when $\Delta\in(\Delta_{\rm c}^{-},\Delta_{\rm c}^{+})$, $
\bra{\Psi_{\rm gs}(\{d_{\boldsymbol{n}}\},\{\pm\frac{\pi}{2},\mp\frac{\pi}{2}\})} H_{\mathsf{Z_2BH}}^{U\to\infty}\ket{\Psi_{\rm gs}(\{d_{\boldsymbol{n}}\},\{\pm\frac{\pi}{2},\mp\frac{\pi}{2}\})}
=\bra{\psi_{\rm f}(\{d_{\boldsymbol{n}}\})}H_{\mathsf{BDI}}(t,\mp\alpha)\ket{\psi_{\rm f}(\{d_{\boldsymbol{n}}\})}$, which corresponds to two possible instances of the dimerized H\"{u}ckel model~\eqref{eq:huckel}. The corresponding dimerization parameter  depends on the two possible symmetry-breaking patterns for the case of hard-wall boundary conditions:

{\it (a)} If the $\mathbb{Z}_2$ fields break the translational symmetry by adopting the N\'eel configuration $\ket{{\uparrow\downarrow\uparrow\downarrow\cdots\uparrow\downarrow\uparrow}}$, the hardcore bosons are subjected to $H_{\mathsf{BDI}}(t,-\tilde{\delta})$ with the pattern of dimerized tunnelings $\{t(1+\tilde{\delta}),t(1-\tilde{\delta})\cdots,t(1-\tilde{\delta}),t(1+\tilde{\delta})\}$, where we recall that $\tilde{\delta}>0$. According to our discussion below Eq.~\eqref{eq:winding}, this regime has a negative dimerization parameter $\delta=-\tilde{\delta}<0$, and the groundstate is a trivial  insulator with  a vanishing winding number $\nu=0$.

 {\it (b)} If the $\mathbb{Z}_2$ fields, instead, break the translational symmetry via $\ket{{\downarrow\uparrow\downarrow\uparrow\cdots\downarrow\uparrow\downarrow}}$, the hardcore bosons are subjected to $H_{\mathsf{BDI}}(t,+\tilde{\delta})$, and thus see the pattern of dimerized tunnelings $\{t(1-\tilde{\delta}),t(1+\tilde{\delta})\cdots,t(1+\tilde{\delta}),t(1-\tilde{\delta})\}$. According to our discussion below Eq.~\eqref{eq:winding}, this regime has a positive dimerization parameter $\delta=+\tilde{\delta}>0$, and the half-filled groundstate is a $\mathsf{BDI}$ topological band insulator with  a non-vanishing winding number $\nu=1$ for $\tilde{\delta}<2$. Note that the symmetry-breaking long-range order of the BOW phase~\eqref{eq:BOW_parameter} occurs simultaneously with the symmetry-protected topological invariant $\nu=1$. Moreover, both of these orders develop in the same degree of freedom: the hardcore bosons. Accordingly, our model yields a clear instance of a symmetry-breaking topological insulator~\cite{SBT_fermions_triang_lattice}: a {\it  topological bond-ordered wave} (TBOW). Let us also emphasize that this interplay between spontaneous symmetry breaking and symmetry-protected topological phases is  characteristic  of our model of lattice bosons coupled to Ising fields~\eqref{eq:z2_bhm}, and cannot be accounted for with the dimerised Bose-Hubbard model~\eqref{eq:dBHM} studied in~\cite{ssh_bose_hubbard}. 
 
 Another feature of this non-trivial topological state is the presence of localized edge states for finite system sizes. In the hardcore limit, chiral symmetry guarantees that these edge states are protected against perturbations that respect the symmetry, as long as the gap does not close \cite{hatsugai_topological_edge}. This is the so-called bulk-boundary correspondence, which relates a quantize topological invariant in the bulk of the system with protected edge states at the boundaries \cite{Asboth_2016}.

\paragraph{Quasi-adibiatic regime: fluctuation-induced topological phase transitions.--} We have seen that the Born-Oppenheimer ansatz~\eqref{eq:BO_var_ansatz} allows us to draw a clear analogy with the Peierls transition in the $\beta=0$ limit, and a transparent understanding of symmetry-breaking topological insulators in the $\mathbb{Z}_2$BHM~\eqref{eq:z2_bhm}.   We can extend this analysis to the quasi-adiabatic regime  $\beta\ll t$, expanding thus the analytical understanding of the bosonic Peielrs mechanism presented in~\cite{paper}  to a situation where  the $\mathbb{Z}_2$ field  dynamics  introduce quantum fluctuations that can modify the Peierls mechanism. 

As shown in Fig.~\ref{fig:BO_phase_diagram},  the quantum dynamics of these fields competes against the formation of the bond-ordered density modulations, modifying the static phase boundaries~\eqref{eq:static_critical_lines} that delimit the TBOW phase. In fact, it is possible to get analytical expressions of how these critical lines get deformed by considering   the variational energies~\eqref{eq:var_energy}  of the previous polarized/N\'eel-type phases for small deviations of the angles $\bs \theta$ around the  corresponding values $\bs \theta^\star$. A comparison of these energies leads to the following critical lines
\beq
\label{eq:dynamical_critical_lines}
\Delta_{\rm c}^{\pm}=\frac{4t}{\pi}\left(\tilde{\delta}\pm\left(\mathsf{E}\big(1-\tilde{\delta}^2\big)-1\right)\right)\mp\frac{\pi\beta^2}{4t\tilde{\delta}\mathsf{E}\big(1-\tilde{\delta}^2\big)},
\eeq
which are represented by the dashed white lines of Fig.~\ref{fig:BO_phase_diagram}, and yield a very good approximation of the yellow region enclosing the symmetry-broken TBOW phase. As advanced previously, these critical lines predict that the area of the TBOW phase decreases as $\beta$ increases, and lead to fluctuation-induced topological phase transitions  connecting the TBOW phase to other trivial band insulators as the   $\mathbb{Z}_2$ field dynamics  becomes more relevant.

\subsubsection{Softcore bosons coupled to $\mathbb{Z}_2$ fields}

In the previous subsection, we have presented a Born-Oppenheimer variational treatment of the $\mathbb{Z}_2$BHM~\eqref{eq:z2_bhm} in the limit of hardcore bosons and quasi-adiabatic $\mathbb{Z}_2$ fields.  Our variational ansatz for the groundstate~\eqref{eq:BO_var_ansatz} and low-energy excitations~\eqref{eq:BO_var_ansatz_exc} has allowed us to draw a clear analogy with the  Peierls instability of 1D metals via the fermionization of the hardcore bosons:  at $U\to\infty$, a Fermi surface emerges and an energy gap can be opened. This analogy has allowed us to show that symmetry-breaking quantum phase transitions can take place at various $(\alpha_{\rm c},\Delta_{\rm c})$, delimiting a finite region of a  TBOW for hardcore bosons (see Fig.~\ref{fig:BO_phase_diagram}). The question we would like to address in this subsection is if such a TBOW phase can only be defined  at the singular $``U=\infty"$ point or if, on the contrary, it persists within the physically-relevant regime of finite  Hubbard interactions.  

The  regime  of strong, yet finite,  interactions  can give rise to interesting strongly-correlated behaviour that cannot be accounted for by  considering solely the $``U=\infty"$ point. For instance, for the Fermi-Hubbard model  close to half-filling, whereas the groundstate is a fully-polarized ferromagnet~\cite{nagaoka_ferromagnetism,nagaoka_finite_densities,nagaoka_dmrg} for  infinite interactions, the regime of 
finite repulsion $0<t/U\ll1$ gives rise to   anti-ferromagnetic super-exchange interactions~\cite{superexchange_anderson,t_U_expansion} that are believed to play a key role in  high-temperature superconductivity~\cite{high_tc_anderson}. Similar super-exchange interactions  also appear in the strongly-interacting limit of two-component Bose-Hubbard models~\cite{super_exchange_bhm}. Such spin-spin interactions are absent in the single-component Bose-Hubbard model, where one obtains density-density interactions between bosons at nearest-neighbouring sites, as well as density-dependent correlated tunnelings~\cite{t_U_bose_hubbard}. 

We note that  in our $\mathbb{Z}_2$BHM~\eqref{eq:z2_bhm}, despite consisting of single-component bosons, the  strongly-interacting limit can be  richer as the virtual tunnelings  are dressed by the corresponding $\mathbb{Z}_2$ link fields. Therefore, in addition to the aforementioned effects, an effective spin-spin interaction between the spins at neighboring links  can also appear as corrections to the $U\to\infty$ limit are studied. To leading order in a  $0<{\rm max}\{t/U,\alpha/U\}\ll1$ expansion, we find that the $\mathbb{Z}_2$BHM~\eqref{eq:z2_bhm} can be expressed as $H_{\mathsf{Z}_2\mathsf{BH}}\approx H_{\mathsf{Z}_2\mathsf{BH}}^{U\to\infty}+\Delta H$, 
where the leading corrections are
\begin{widetext}
\beq
\label{eq:perturbation}
\Delta H=-\frac{4t^2}{U}\sum_i\left(1+2\tilde{\delta}\sigma_{i,i+1}^z+\tilde{\delta}^2\right)n_in_{i+1}+\frac{2t^2}{U}\sum_i\left(1+\tilde{\delta}\big(\sigma_{i,i+1}^z+\sigma_{i+1,i+2}^z\big)+\tilde{\delta}^2\sigma_{i,i+1}^z\sigma_{i+1,i+2}^z\right)\!\left({c}^\dagger_i n^{\vphantom{\dagger}}_{i+1}{c}^{\vphantom{\dagger}}_{i+2}+{\rm H.c.}\right),
\eeq
\end{widetext}
and we have introduced the density operators $n_i^{\vphantom{\dagger}}={c}^{{\dagger}}_{i}{c}^{\vphantom{\dagger}}_{i}$. The first term describes the second-order process where the boson virtually tunnels back and forth to a neighboring  occupied site,  giving rise to a virtual double occupancy and to density-density couplings, which cannot be accounted in the hardcore-boson limit  $U\to\infty$. Note that this virtual tunneling is mediated by the link $\mathbb{Z}_2$ field, which can thus modify the strength of the density-density coupling depending on the background configuration of the spins. The second term describes the second-order process where a boson virtually tunnels between two sites apart via an intermediate occupied site, giving rise to a density-dependent correlated tunneling. Note again, that this virtual tunneling is dressed by the   link $\mathbb{Z}_2$ fields, and its strength can depend on their configuration, including spin-spin correlations. From a different perspective, these mediated virtual tunnelings give rise to an effective coupling between neighboring link spins, as announced above.

As discussed in the hardcore-boson limit below Eq.~\eqref{eq:BOW_parameter}, the Born-Oppenheimer approximation allows us to extract an effective Hamiltonian for the bosonic sector, which depends on the $\mathbb{Z}_2$ field configuration in the ground state. Due to the finite $U$ corrections (\ref{eq:perturbation}), the effective Hamiltonian for softcore bosons contains interaction terms and cannot be reduced to a single-particle model. Moreover, these terms break the chiral symmetry of the system, which protects the TBOW phase in the limit $U \rightarrow \infty$ (see our discussion below Eq.~\eqref{eq:winding}). Nevertheless, the effective Hamiltonian with corrections~\eqref{eq:perturbation} is still invariant under a bond-centered inversion symmetry~\cite{ssh_bose_hubbard}. Therefore, in analogy  to the Bose-Hubbard model with dimerized fixed tunnelings (\ref{eq:dBHM}), the ground state of the effective Hamiltonian shall correspond to a SPT phase protected by inversion symmetry as long as the $\mathbb{Z}_2$ field displays a N\'eel-type anti-ferromagnetic ordering and the gap remains open. Following a topological argument, as long as the gap remains open and the symmetry is present, the many-body generalization~\cite{berry_phase_review} of the topological invariant~\eqref{eq:zak_phase} will be quantized to the same value as in the hardcore limit, even in softcore regimes away from  the singular $``U=\infty"$ point. In the next section, we will confirm this prediction by computing numerically the topological invariant in the strongly-correlated bosonic regime.

In order to show that the energy gap remains open in the softcore regime, we can explicitly calculate how the variational excitation energies~\eqref{eq:minimization} get modified due to the perturbations in Eq.~\eqref{eq:perturbation}.  Our variational ansatz for the excitations allows us to go beyond standard mean-field theory, and obtain corrections to the excitation branches of Eq.~\eqref{eq:qpart_energies}, giving rise to dispersive spin-wave-type excitations, or coupled spin-boson quasi-particles. For the many-body gap, as discussed in Appendix~\ref{sec:app:details},  the TBOW energy gap~\eqref{eq:gap_opening} is shifted  to
\beq
\label{eq:gap_closing}
\begin{split}
\Delta\epsilon&\approx 2t(\bs \theta^\star)|\delta(\bs \theta^\star)|\\
&-\frac{4t^2}{U}\left(1+\frac{1}{\pi}\mathsf{E}(1-\tilde{\delta}^2)\right)\left(1+\tilde{\delta}\left(\sin\theta_A^\star+\sin\theta_B^\star\right)\right),
\end{split}
\eeq
 to leading order in $\tilde{\delta}\ll1$. At the level of our Born-Oppenheimer ansatz, we see that the energy gap remains finite, such that the TBOW phase extends to the soft-core regime. Moreover,
one can also see that the TBOW gap decreases as the interactions are lowered. This trend can be also be understood  from the following alternative perspective. As discussed above,  the virtual tunnelings give rise to an effective coupling between neighbouring link spins. Since fermions are much faster than the spins, and the $t/U$ corrections~\eqref{eq:perturbation} are assumed to be small, the value of this coupling can be approximated using the fermionic unperturbed groundstate  in Eq.~\eqref{eq:BO_var_ansatz}
 $\bra{\psi_{\rm f}(\{d_{\boldsymbol{n}}\})} \left({c}^\dagger_i n^{\vphantom{\dagger}}_{i+1}{c}^{\vphantom{\dagger}}_{i+2}+{\rm H.c.}\right) \ket{\psi_{\rm f}(\{d_{\boldsymbol{n}}\})}$.
We can evaluate this expectation value by applying Wick's theorem, as the variational ansatz is built with free spinless fermions. This calculation shows that effective spin-spin interaction is ferromagnetic, which would compete against the N\'eel-type order  of the $\mathbb{Z}_2$ fields,   making the TBOW phase less stable (i.e. lowering the corresponding energy gap). 

Although it cannot be captured by our  variational approach, this tendency opens the possibility that  the energy gap  closes for sufficiently small interactions, such that a quantum phase transition to a non-topological phase takes place. Accordingly, the TBOW phase may not be adiabatically connected to a bosonic non-interacting SPT phase, and one could claim that  it is  an instance of an interaction-induced symmetry-breaking topological insulator.
 In order to explore this possibility further,  and to benchmark the qualitative correctness of our predictions based on the Born-Oppenheimer variational approach, we now move onto a quasi-exact numerical  method based on the density-matrix renormalization group (DMRG). 

\section{Topological bond-order wave: a DMRG approach}
\label{sec:dmrg}

In this section, we use the density matrix renormalization group algorithm \cite{tenpy} to study the properties of the TBOW phase. We benchmark the previous variational results by exploring the  strongly-correlated regime of finite Hubbard interactions $U$, and dynamical $\mathbb{Z}_2$ fields $\beta>0$. We start by giving compelling evidence to show that the BOW phase is indeed a SPT phase protected by a bond-centered inversion symmetry. To this end, we use both the entanglement spectrum and a local topological invariant to characterize the topological nature of the phase. We also show the presence of many-body localized edge states with a fractional particle number. Finally, we study the transition from the TBOW phase to a non-topological quasi-superfluid (qSF) phase (see Fig.~\ref{fig:scheme}{\bf (b-c)}) as the Hubbard interactions are lowered, and present a phase diagram of the model. Our numerical results clearly show the need of both strong interactions and symmetry breaking in  order to stabilize the TBOW phase, which cannot be connected to a non-interacting topological insulator.

\subsection{Symmetry-breaking order parameters}

\begin{figure}[t]
  \centering
  \includegraphics[width=0.85\linewidth]{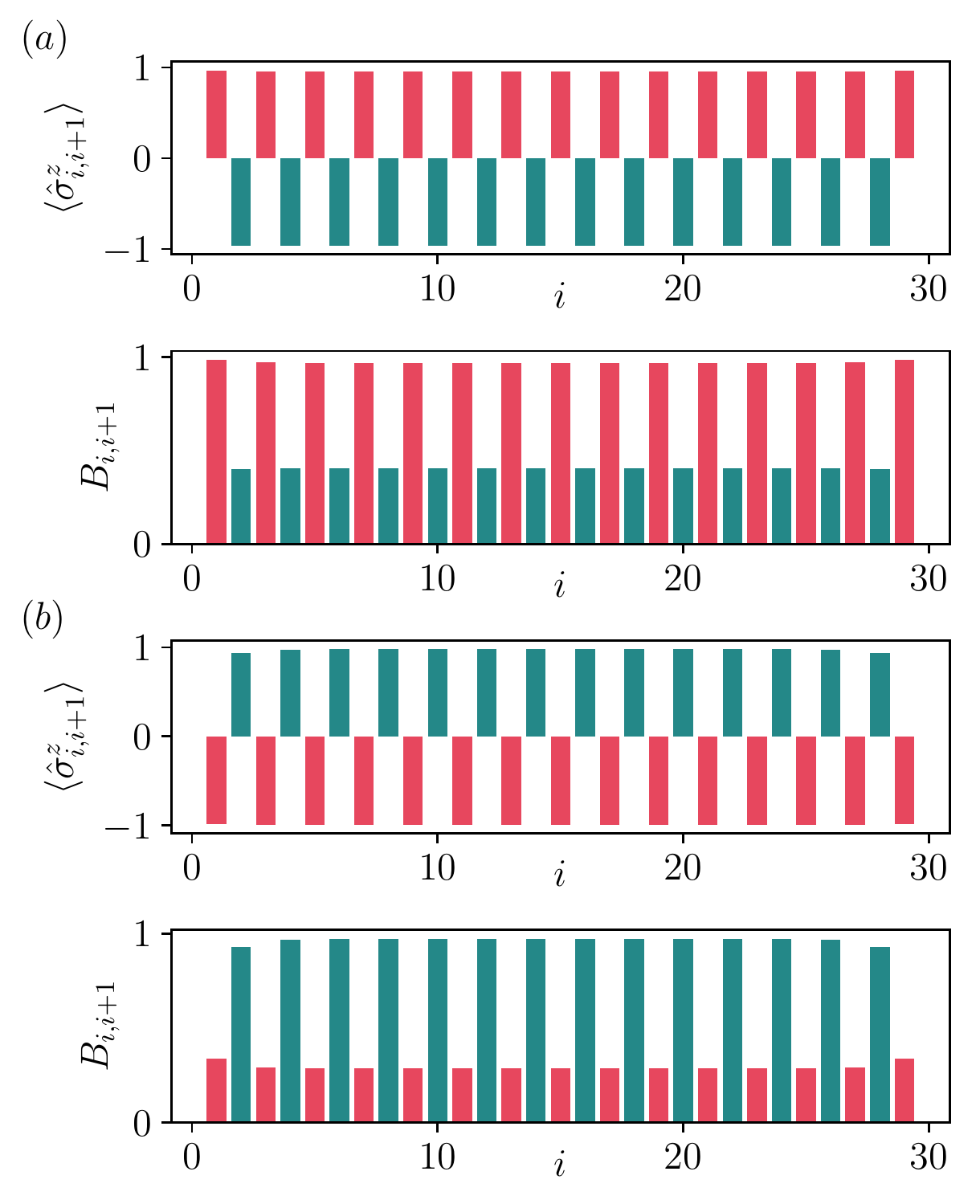}
\caption{\label{fig:bow_op}{\bf Symmetry-breaking order parameters:} Spin pattern $\langle\hat{\sigma}^z_{i,i+1}\rangle$ and particle density $\mathsf{B}_{i,i+1}$ on the bonds of the lattice for a half-filled system. Different colors depict even and odd  bonds. {\bf (a)} Symmetry-broken sector with the $\mathbb{Z}_2$ fields in the approximate N\'eel configuration  $\ket{{\uparrow\downarrow\uparrow\downarrow\cdots\uparrow\downarrow\uparrow}}$, which leads to a period-two strong-weak modulation of the bond density (see the scheme in Fig.~\ref{fig:scheme}{\bf (b)}). {\bf (b)} Symmetry-broken sector with the $\mathbb{Z}_2$ fields in the approximate N\'eel configuration  $\ket{{\downarrow\uparrow\downarrow\uparrow\cdots\downarrow\uparrow\downarrow}}$, which leads to a period-two weak-strong  modulation of the bond density.}
\end{figure}

According to our variational ansatz in the hardcore limit, and the discussion below Eq.~\eqref{eq:BOW_parameter}, there should be a finite region of parameter space hosting a TBOW phase (see Fig.~\ref{fig:BO_phase_diagram}). Figure~\ref{fig:bow_op} shows the numerical DMRG  results for the order parameters characterizing the BOW phase of the $\mathbb{Z}_2$BHM~\eqref{eq:z2_bhm}, focusing on strongly-correlated bosons coupled to quasi-adiabatic $\mathbb{Z}_2$ fields  (i.e. $U = 20t$, and $\beta = 0.01t$). For these results, and for the rest of the paper, we use a chain of $L = 30$ sites and a bond dimension $D = 100$, and fix the rest of the Hamiltonian parameters to $\alpha = 0.5t$, $\Delta = 0.8t$. As described in the previous section, the BOW phase is partially characterized by the spontaneous breaking of translational invariance, which is captured by the alternation of the $\mathbb{Z}_2$ magnetization $\langle\sigma_{i,i+1}^z\rangle$, and the modulation of the bond density $B_{i,i+1}$. Moreover, this phase is gapped and incompressible, as it was shown in \cite{paper}. Long-range order develops in the system after the symmetry breaking, whereby the unit cell of the system is doubled. The two possible symmetry-broken ground-states of   Figs.~\ref{fig:bow_op} {\bf (a)} and {\bf (b)} are completely degenerate in the thermodynamic limit, and they can be connected by a one-site lattice translation. For finite lattice sizes, they differ at the edge of the system. Let us remark, however,  that these bonding/antibonding order parameters do not suffice to capture all the physics  of the BOW phase, as they do not account for the  topological features that make the two symmetry-broken sectors fundamentally different from each other.

\subsection{Topological characterization of the TBOW} 
\subsubsection{Entanglement spectrum}

We first explore numerically the entanglement properties of the ground-state. In particular, we compute the entanglement spectrum~\cite{li_haldane}. We define a bipartition of the system, and write the ground-state as $\ket{\psi_{\rm gs}} = \sum_n \tilde{\lambda}_n \ket{\psi_n}_\mathfrak{L} \otimes \ket{\psi_n}_\mathfrak{R}$, where $\mathfrak{L}$ and $\mathfrak{R}$ are the two subsystems, and $\{\tilde{\lambda}_n\}$ are the corresponding Schmidt coefficients. The entanglement spectrum is defined as the set of all the Schmidt coefficients in logarithmic scale $\epsilon_n = - 2 \, \text{log}(\tilde{\lambda}_n)$. It has been established that the entanglement spectrum is degenerate for symmetry-protected topological phases \cite{entanglement_spectrum}. In particular, this degeneracy is robust against perturbations that respect the symmetry as long as the many-body gap of the system is open.

\begin{figure}[t]
  \centering
  \includegraphics[width=1.0\linewidth]{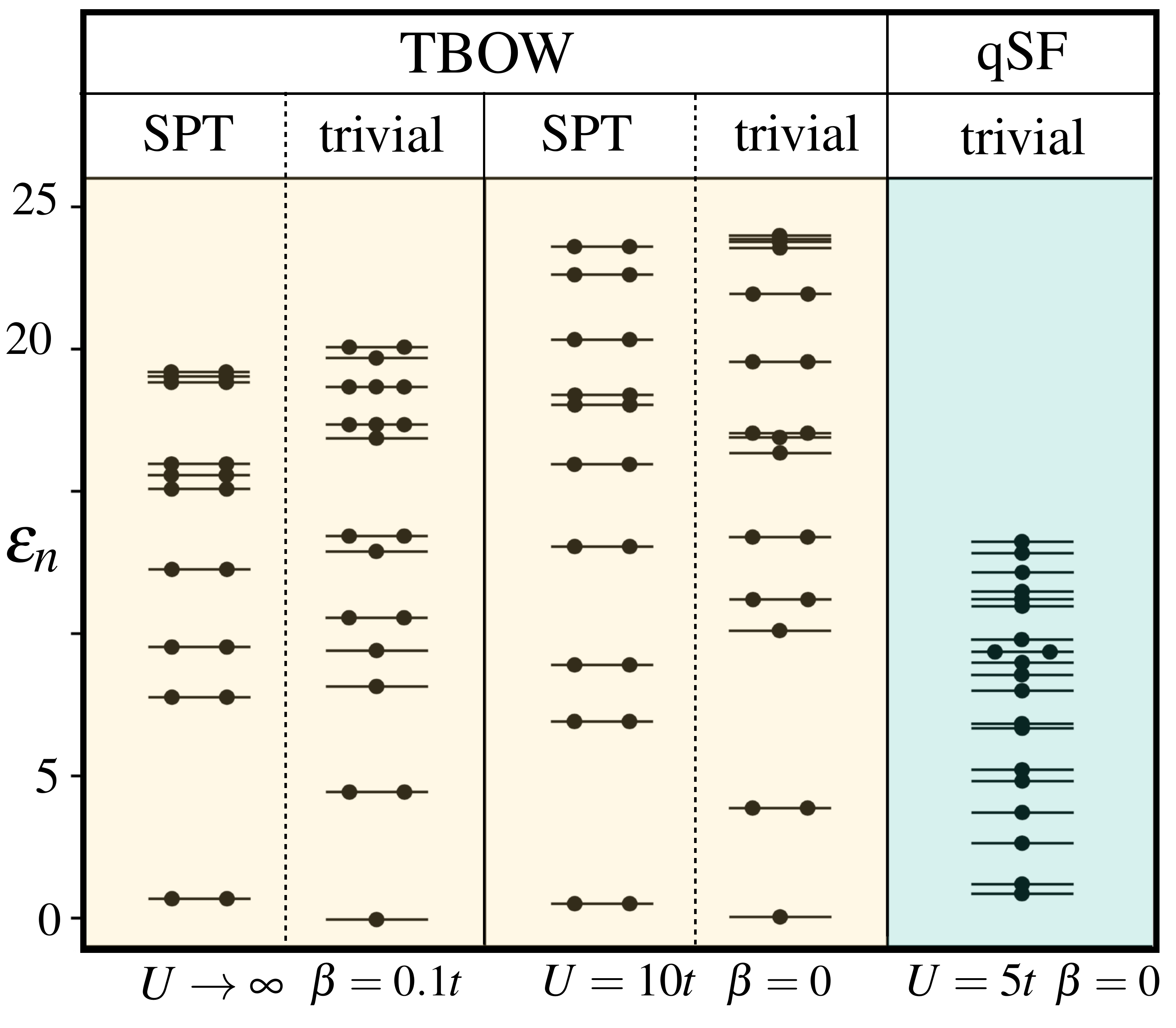}
\caption{\label{fig:entanglement} {\bf Entanglement spectrum degeneracies:} Lower $20$ eigenvalues of the entanglement spectrum for different states in the TBOW and the quasi-SF phases. For the former, we show the spectrum in the two different symmetry-breaking sectors. We can see how the spectrum is double degenerate in one case, which corresponds to the non-trivial topological sector (see our discussion below Eq.~\eqref{eq:BOW_parameter}). We observe how the degeneracies survive in two strongly-correlated regimes: for hardcore bosons ($U = \infty$) coupled to dynamical $\mathbb{Z}_2$ fields ($\beta = 0.1t$), and for soft-core bosons ($U = 10t$) coupled to static $\mathbb{Z}_2$ fields  ($\beta = 0$). This degeneracy is lost for small enough interactions in the non-topological quasi-SF phase.}
\end{figure}

In Fig.~\ref{fig:entanglement} (left panel), we present the entanglement spectrum for the BOW phase in the hardcore-boson limit. We consider a bipartition at the middle of the chain, and explore the  two possible degenerate ground-state configurations that appear as a consequence of the spontaneous breaking of  translational symmetry. As discussed below Eq.~\eqref{eq:BOW_parameter}, we expect that the weak-strong bond-density modulation (cf. Fig.~\ref{fig:bow_op}{\bf (b)}) due to the  symmetry-broken background of $\mathbb{Z}_2$ fields gives rise to a SPT phase. As follows from Fig.~\ref{fig:entanglement} (left panel), the entanglement spectrum is two-fold degenerate for one of the ground-states, while it clearly lacks an exact two-fold degeneracy for the other configuration.  These numerical results provide a  clear signature of the topological nature of the BOW phase, and confirm the  scenario of the interplay between symmetry breaking and SPT phase of the $\mathbb{Z}_2$BHM~\eqref{eq:z2_bhm} predicted by the Born-Oppenheimer variational approach.

In the central panel of  Fig.~\ref{fig:entanglement}, we show the entanglement spectrum for strongly-correlated bosons on a static lattice ($U = 10t$, $\beta = 0$), thus exploring the departure from the hardcore constraint. As discussed below Eq.~\eqref{eq:gap_closing}, for strong yet finite interactions, we expect that the energy gap is finite, and that the  TBOW phase persists as one lowers the interactions. This expectation is supported by our numerical results, which again display a clear two-fold degeneracy of the entanglement spectrum in one of the symmetry-broken groundstates. Let us finally note that, for sufficiently weak interactions ($U = 5t$, $\beta = 0$), the degeneracy of the spectrum is completely lost for the single ground-state of the system (see the right panel of Fig.~\ref{fig:entanglement}).  This non-topological phase for the weakly-interacting bosons corresponds to the quasi-superfluid (SF) schematically depicted in Fig.~\ref{fig:scheme}{\bf (b-c)}, and will be discussed in more detail in Sec.~\ref{sect:phase_structure} below. The latter facts again support our claim that this strongly-correlated TBOW phase has an interaction-induced nature, as the topological features are completely absent in the weakly-interacting regime.

Let us finally emphasize that our entanglement spectrum analysis away from the hardcore limit has been restricted to static $\mathbb{Z}_2$ fields, which is necessary as the calculation of the entanglement spectrum requires a bipartition of the system that respects the protecting symmetry of the topological phase. In the case of  hardcore bosons, the system is protected by chiral symmetry and the half-chain bipartition respects that symmetry. On the other hand, for finite interactions,  the phase is instead protected by inversion symmetry (see our discussion below Eq.~\eqref{eq:perturbation}). For our DMRG implementation, the presence of the $\mathbb{Z}_2$ fields does not allow us to cut the system into two halves in such a way that the two parts respect the bond-centred inversion symmetry. This is only possible for static spins ($\beta = 0$), since they form a product state and do not contribute to the entanglement properties of the system. This particularity of our model prevents us from using the entanglement spectrum to characterize the topological properties in the regime of strongly-correlated bosons coupled to dynamical fields ($U$ finite, $\beta>0$), although the topological nature of the TBOW phase must also be preserved as the transverse field $\beta$ is slightly increased (see Fig.~\ref{fig:BO_phase_diagram}). For this reason, we introduce in the next section a robust topological invariant that yields an alternative route to characterize  the topological properties of the BOW phase in any parameter regime.

\subsubsection{Local Berry phase}

We now characterize the topology with the help of the local Berry phase introduced by Hatsugai~\cite{hatsugai_local}. It is a topological invariant that serves as a local ``order parameter" to distinguish symmetry-protected topological phases in the presence of interactions. Considering a periodic Hamiltonian ${H}(\lambda)$, which depends on an external parameter $\lambda \in [\lambda_0,\lambda_{\rm f}]$ through an adiabatic cyclic evolution ${H}(\lambda_0) = {H}(\lambda_{\rm f})$.  As shown in~\cite{hatsugai_local}, if there exists an antiunitary operator $\Theta=K U_\Theta$,  where $U_\Theta$ is unitary and $K$ is the complex conjugation, which commutes with ${H}(\lambda)$,  the Berry phase~\cite{berry_phase} acquired by the groundstate $ \ket{\psi_\lambda} $ during the parallel transport on a  loop $C$ with $\lambda_{\rm f}=\lambda_0$,
\begin{equation}
\label{eq:zak_rs}
\gamma_C=\ii\oint_C\mathrm{d} \lambda\, \langle \psi_\lambda \vert \partial_\lambda \psi_\lambda \rangle {\rm mod}2\pi,
\end{equation} 
is quantized to finite values $0$ and $\pi$. Let us note that, for non-interacting systems, considering the quasi-momentum as the cyclic parameter $\lambda=k$ as one traverses the Brillouin zone,  Eq.~\eqref{eq:zak_rs}
 reduces to the previously-introduced Zak's phase $\gamma_{\rm C}=\varphi_{\rm Zak}$~\eqref{eq:zak_phase}. However, this free-fermion topological invariant cannot be directly applied to interacting systems. Alternatively,  we shall use Eq.~\eqref{eq:zak_rs} with a different adiabatic parameter that introduces the notion of locality, and allows us to generalize the topological invariant to a many-body scenario.

This quantity is topological in the sense that it cannot change without  closing the gap, as long as the corresponding symmetry is preserved. One can add, in particular, a local perturbation to the initial Hamiltonian~\eqref{eq:z2_bhm}, like a local twist in one of the hopping strengths $t\rightarrow t \lambda=te^{i\theta}$, which does not close the gap of the BOW phase.  Note that this is similar to the use of twisted boundary conditions to calculate the Zak phase in the presence of interactions~\cite{berry_phase_review}. However, as noticed by Hatsugai~\cite{hatsugai_local}, this perturbation should not be necessarily put on the edges of the system,  but can be placed on any bond as long as it preserves the symmetry. This choice is appropriate in our case, since we can add local perturbations that respect inversion symmetry, and constitute therefore a local measure in the bulk of the system. Moreover, it only depends on quantities that decay exponentially and, thus, this measurement is  valid not only for periodic, but also for open hard-wall boundary conditions, as long as the perturbation is not applied too close to the edge of the system.

\begin{figure}[t]
  \centering
  \includegraphics[width=0.85\linewidth]{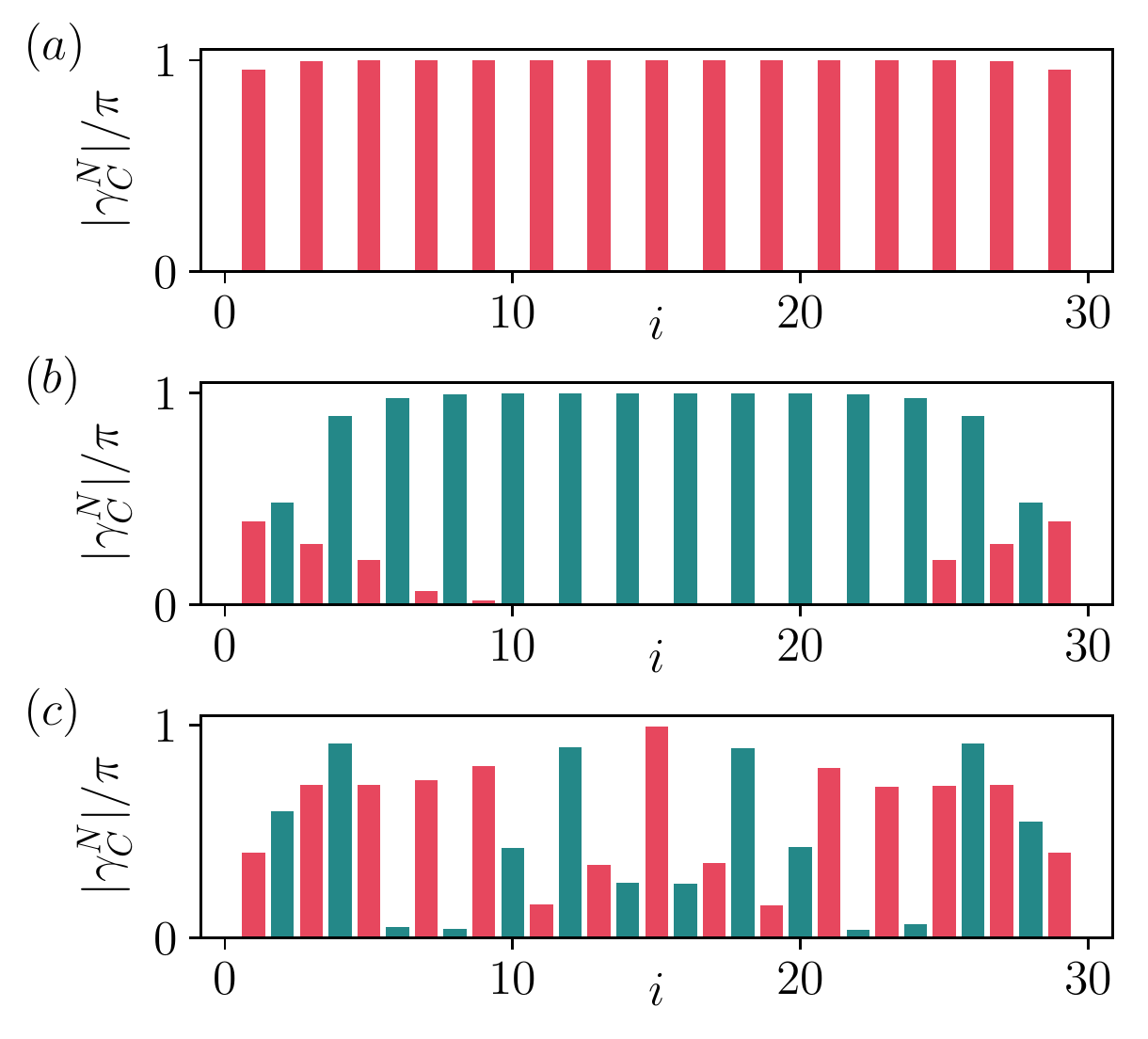}
\caption{\label{fig:berry} {\bf Local Berry phase quantization:}  local Berry phase~\eqref{eq:discretized_berry} calculated for each bond $\gamma^N_C (i,i+1)$, such that  even and odd bonds are depicted in different colors.  We use a value of $N = 5$ for three different ground states: {\bf (a)} and {\bf (b)} correspond to the trivial and topological  symmetry-broken sectors of the  BOW phase,  respectively, for $U = 20t$ and $\beta = 0.02t$. In the bulk, the phases are quantized to values $0$ and $\pi$, alternating between even and odd bonds, and for each bond between the two sectors.  ({\bf c)} Configuration for a state in the quasi-superfluid phase, with $U = 5t$ and $\beta = 0.02$, where the translational symmetry is not broken. In this case, the phases are not quantized since the phase is not a SPT phase.}
\end{figure}

\begin{figure}[t]
  \centering
  \includegraphics[width=0.9\linewidth]{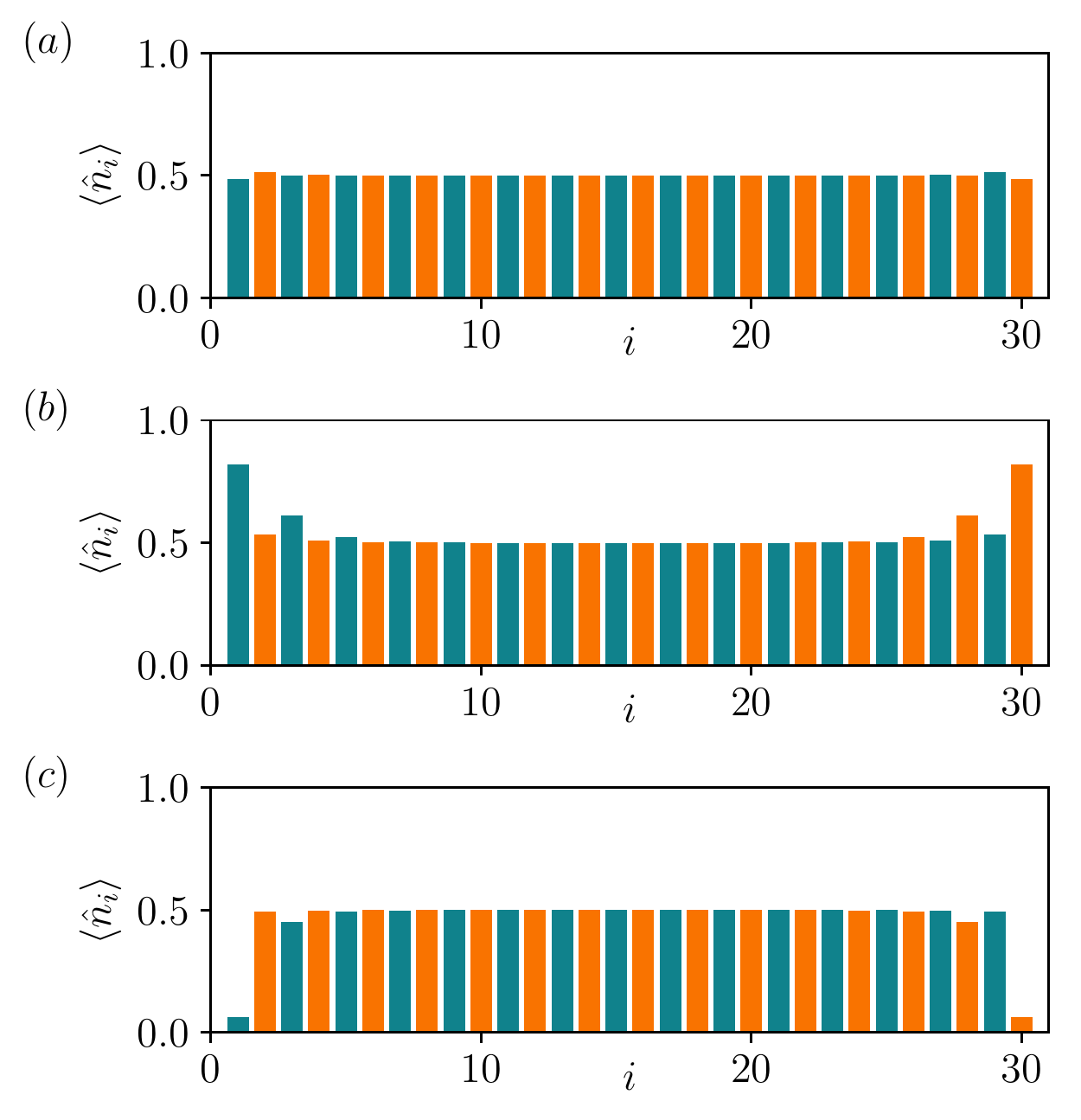}
\caption{\label{fig:bow_edges} {\bf Many-body edge states:} Real-space configuration of the bosonic occupation number $\langle\hat{n}_i\rangle$, using different colors for even and odd sites. {\bf (a)} The trivial topological sector of the BOW phase is characterized by the absence of edge states in the bosonic sector. The non-trivial sector is represented in {\bf (b)} and {\bf (c)} for one extra particle and one extra hole, respectively. We can see how the extra particle (resp. hole) generates two edge states, each one carrying a fractional particle number of $+ 1 / 2$ (resp. $- 1 / 2$). }
\end{figure}

In practice, the integral of Eq.~\eqref{eq:zak_rs} can be challenging to compute as the integrand is gauge dependent (the integral on the loop, however, is gauge invariant), requiring a numerical gauge fixing at each discretized point of the loop. Alternatively, we compute here the integral using a Wilson loop formulation, which is gauge invariant and avoids the gauge fixing problem \cite{hatsugai_local,hatsugai_algorithm}, namely
\begin{equation}
\label{eq:discretized_berry}
\gamma^N_C = \mathrm{Arg} \prod_{n = 0}^{N-1} \langle \widetilde{\psi}_n | \widetilde{\psi}_{n + 1} \rangle
\end{equation}
where $| \widetilde{\psi}_n \rangle = \ket{\psi_{\lambda_n}}\braket{\psi_{\lambda_n}}{\phi}$ is the projection of a reference state $\ket{\phi}$ onto the adiabatic groundstate, with $\lambda_0,\,\lambda_1,...,\,\lambda_N = \lambda_0$ being  the $N$ points in which the loop $C$ is discretized. 
The discretized local Berry phase~\eqref{eq:discretized_berry} depends in general on $N$ and on the way the loop is discretized, but not on the reference state as far as it has a finite overlap with the ground-state. However, we note that $\gamma^N_C$ converges rapidly to the local Berry phase (\ref{eq:zak_rs}) in the large $N$ limit.

 We define the local phase $\gamma^N_C (i,i+1)$, corresponding to the bond $(i,i+1)$, by adding a local perturbation to the bare tunneling coefficient, $te^{i2\pi n / N}$, with $n\in \{0,...N\}$. Note that this perturbation preserves the bond-centered inversion symmetry, and thus does not present the limitations of the entanglement spectrum mentioned in the previous section. Therefore, we can use it to explore the topological features of the TBOW phase away from the hardcore boson limit and considering dynamical $\mathbb{Z}_2$ fields in the quasi-adiabatic regime. Figure~\ref{fig:berry} shows the local Berry phases at every bond for the TBOW and the qSF phases.
Figs.~\ref{fig:berry}{\bf (a)} and {\bf (b)} correspond to the two degenerate ground states of the TBOW for $U = 20t$ and $\beta = 0.02t$, where we observe a Berry phase  quantized to values of $0$ and $\pi$ in the bulk of the system. The quantized values alternate for even and odd bonds, and the pattern is reversed for the two ground states, allowing us to assign two different SPT sectors  using the following reasoning.

For finite system sizes, we define the two-site unit cells in such a way that the even bonds\textemdash $(i,i+1)$ with $i$ even\textemdash are intercell, whereas the odd bonds are intracell (i.e. they couple bosons within the same unit cell). If we then focus only on the intercell bonds (green in the figure), we can see how the corresponding local Berry phases are all quantized to $0$ (Fig.~\ref{fig:berry}{\bf (a)}) for the $\mathbb{Z}_2$-field configuration of Fig.~\ref{fig:bow_op}{\bf (a)} adiabatically connected to $\ket{{\uparrow\downarrow\uparrow\downarrow\cdots\uparrow\downarrow\uparrow}}$, which is in agreement with  our variational ansatz that predicted a trivial band insulator for such symmetry-breaking pattern. On the other hand, the  local Berry phases for intercell bonds
for the configuration (Fig.~\ref{fig:bow_op}{\bf (b)}) adiabatically connected to $\ket{{\downarrow\uparrow\downarrow\uparrow\cdots\downarrow\uparrow\downarrow}}$ are all quantized to $\pi$ (Fig.~\ref{fig:berry}{\bf (b)}), which again is in accordance with our variational ansatz predicting a inversion-symmetric SPT phase. Using this convention, the latter can be regarded as the non-trivial topological configuration, and the value of the local Berry phase connects to  that of the Zak phase obtained in the previous section for hardcore bosons in the Born-Oppenheimer approximation. These results confirm our previous expectation that the TBOW phase extends to the softcore regime and for dynamical $\mathbb{Z}_2$ fields, and is characterized by a many-body generalization of Eq.~\eqref{eq:zak_phase} with the same quantized value.
 For $U = 5t$, the groundstate is in a non-topological qSF phase, and the Berry phase does not show a  quantized pattern.  We will show now that this ground state support many-body edge states with fractional particle number.

\subsubsection{Many-body edge states and fractionalization}

For a system with boundaries, an alternative signature of the topological nature of the TBOW phase is the presence of localized edge states, which lie in the middle of the gap for 1D SPT phases with chiral symmetry~\cite{hatsugai_topological_edge}. These edge states are topologically robust against perturbations that respect the symmetry and do not close the gap. Let us note that this bulk-boundary correspondence does not always hold for generic SPT phases, since the presence of edge states might not be guaranteed even if the bulk presents non-trivial topological properties, as is the case of phases protected by inversion symmetry~\cite{entanglement_inversion, topological_insulators_inversion}. In some of these cases, however, localized edge states can be observed in the spectrum as remnants of the protected edge states of a  extended two-dimensional system~\cite{hatsugai_chern}. As shown below, this is precisely the situation for the $\mathbb{Z}_2$BHM~\eqref{eq:z2_bhm}.

Figure~\ref{fig:bow_edges} shows the real-space density configuration of bosons for the two degenerate ground state configurations of the  symmetry-broken BOW phase for finite Hubbard interactions ($U = 20t$, $\beta = 0.02t$). In the topologically-trivial configuration, which is characterized by the long-range order displayed in Fig.~\ref{fig:bow_op}{\bf (a)}, which leads to a period-two strong-weak alternation of the bonds,  we do not observe any localized edge states (cf. Fig.~\ref{fig:bow_edges}{\bf (a)}). On the contrary, for the long-range order characterizing the other symmetry-broken sector in Fig.~\ref{fig:bow_op}{\bf (b)}, which leads to a period-two weak-strong alternation of the bonds, we see localized peaks or drops in the occupation number when we either add {\bf (b)} or subtract {\bf (c)}, respectively, one particle above or below half filling. 

These many-body edge states possess a fractional particle number of $\pm 1 / 2$, which constitutes a bosonic analogue of the predicted charge fractionalization in fermionic quantum field theories~\cite{PhysRevD.13.3398}. In particular, the occupation number $\langle \hat{n}_i \rangle$, which is equal to $1/2$ in the bulk, differs at the edges for the two states. Fractionalization implies that an extra particle or hole is ``divided" into two separate quasiparticles, each carrying half of the particle number. These quasiparticles are localized in different parts of the system and are independent of each other, although they can only be created/annihilated in pairs. The latter can be formed by two fractional $+1/2$ particles, two fractional $+1/2$ holes, or one of each.

As already mentioned, the bulk-boundary correspondence only guarantees the presence of protected edge states in SPT phases protected by chiral symmetry. This is the case for hardcore bosons, where we find protected localized states at the boundaries of the system. However, our DMRG results show that these states are still present for finite Hubbard interactions, even if the protected symmetry is changed from chiral symmetry to  a bond-centered inversion symmetry. Although we can not guarantee the protection of these states, their origin can be understood if we extend the chain to a two-dimensional system, where the bulk-boundary correspondence is restored, and the topological bulk guarantees the existence of one-dimensional conducting states at the boundaries~\cite{hatsugai_chern}.

With the help of these three observables, we have characterized the topological nature of the BOW phase. In particular, using both the entanglement spectrum and the local Berry phase \eqref{eq:discretized_berry}, we proved that one of the two degenerate symmetry-broken states of the BOW phase has a non-trivial bulk topology. Moreover, this topological property persists for finite Hubbard interactions and dynamical $\mathbb{Z}_2$ fields. These numerical evidences confirm the qualitative predictions of the Born-Oppenheimer approximation, and can also be used to explore regimes that lie beyond the applicability of the variational ansatz. Finally, we discussed the presence of many-body edge states in the TBOW phase. All these signatures allow us to regard the TBOW phase as an interaction-induced symmetry-breaking topological insulator protected by a bond-centered inversion symmetry.

\subsection{Interaction-induced nature of the TBOW}
\label{sect:phase_structure}

In this section, we discuss the importance of strong correlations for the existence of the TBOW phase. Using the Born-Oppenheimer approximation, we were able to calculate the single-particle gap in the hardcore boson limit (\ref{eq:gap_opening}), and show that  it gets reduced if we introduce corrections~\eqref{eq:perturbation} for large but finite Hubbard interactions (\ref{eq:gap_closing}). This result suggested  the existence of a phase transition for small enough values of $U$, such that the TBOW phase cannot be adiabatically connected to a non-inetracting SPT phase. Moreover, in the previous section we showed how the signatures of non-trivial topological properties \textemdash such as the degeneracies of the entanglement spectrum (Fig.~\ref{fig:entanglement}) and the quantization of the local Berry phase (Fig.~\ref{fig:berry})\textemdash disappear for small interactions, where one expects the ground state to be in a non-topological qSF phase. In this section, we explore this conjecture and show that, indeed, the TBOW phase can be considered as an interaction-induced SPT phase as one starts from a qSF, and crosses a quantum critical point by increasing the Hubbard interactions.

\begin{figure}[t]
  \centering
  \includegraphics[width=1.0\linewidth]{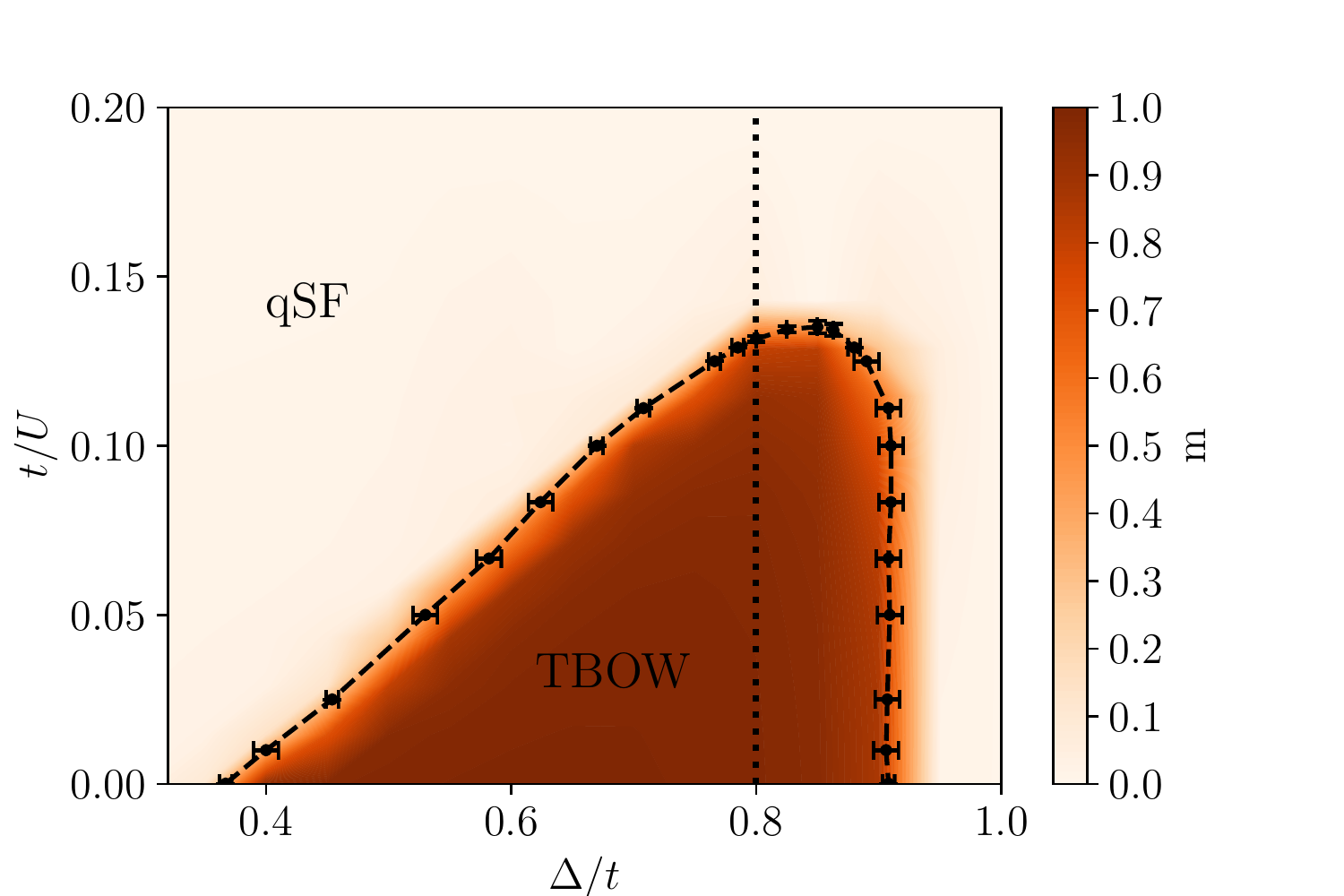}
\caption{\label{fig:phase_diagram}\textbf{Phase diagram:} Phase diagram of the Hamiltonian (\ref{eq:z2_bhm}) in terms of the parameters $\Delta/t$ and $t/U$ for the half-filled case using DMRG. The rest of the parameters are fixed to $\alpha = 0.5t$, $\beta = 0.02t$. The staggered magnetization $m$ for a system size $L = 60$ is represented by the color plot: it has a non-zero value in the TBOW phase and goes to zero in the qSF, allowing the distinction between these two phases. The black dots (with the corresponding error bars) represent the critical points in the thermodynamic limit obtained by a finite-size scaling of $m$ (see Fig. \ref{fig:U_critical}), and the dashed line connecting them is drawn to guide the eye. The dotted vertical line corresponds to the transition for $\Delta = 0.80$ represented in Fig.~\ref{fig:U_critical}. For small enough values of the interaction strength, the ground state of the system is in a qSF phase for any value of $\Delta$. This supports our claim that the TBOW phase is an interaction-induced symmetry-breaking topological insulator.}
\end{figure}

To support this claim, we present in Fig.~\ref{fig:phase_diagram} the phase diagram of the model at half filling in terms of $\Delta/t$ and $t/U$ using DMRG. The color plot represent the staggered magnetization,
\begin{equation}
m = \frac{1}{L}\sum_i (-1)^i\langle\hat{\sigma}^z_{i, i+1}\rangle,
\end{equation}
for a system size of $L = 60$. This order parameter allows one to distinguish between the TBOW and the qSF phase. The phase diagram also shows the critical line obtained in the thermodynamic limit, separating the TBOW and the qSF phase for small enough values of $U$. 

\begin{figure}[t]
  \centering
  \includegraphics[width=1.0\linewidth]{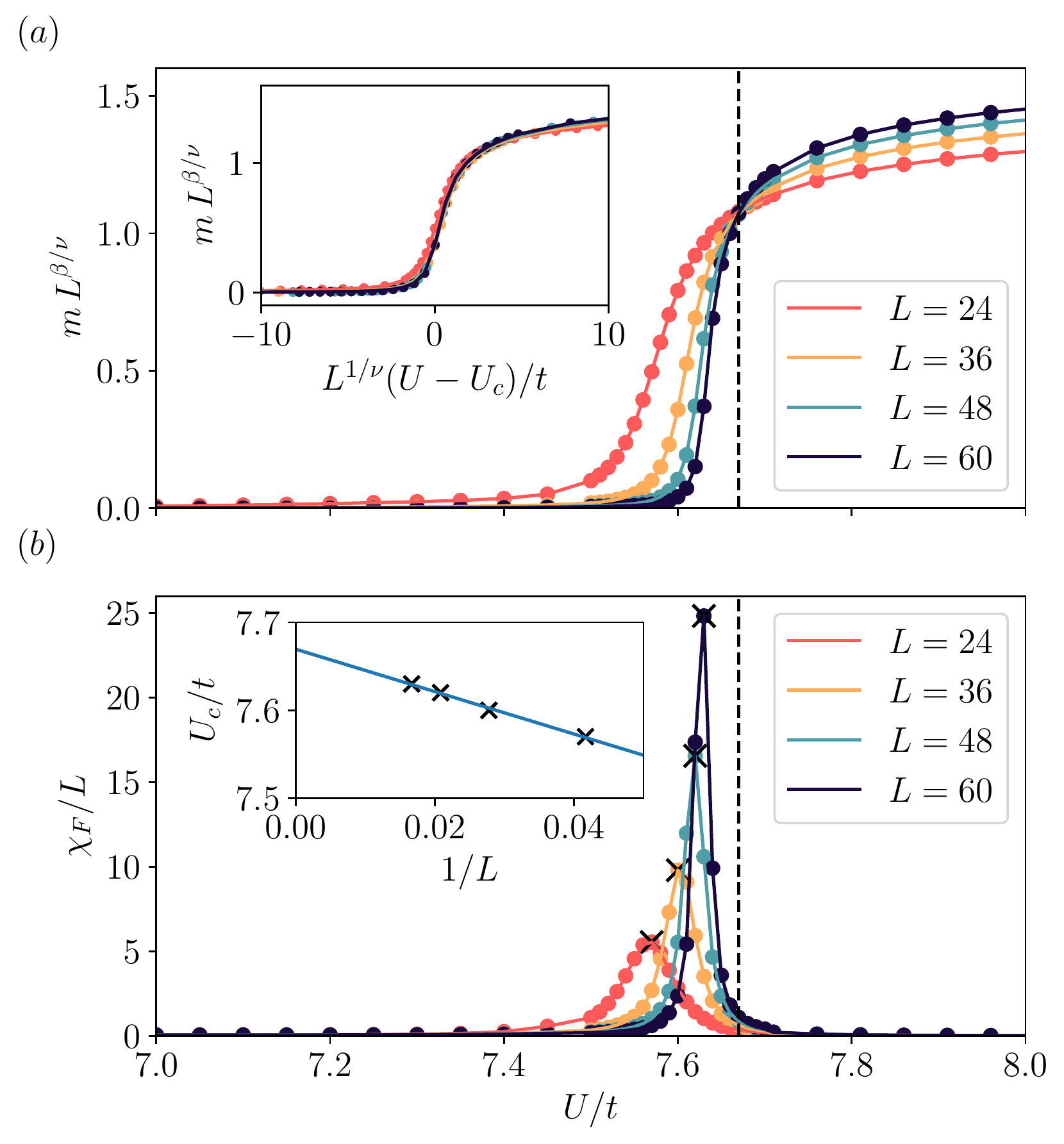}
\caption{\label{fig:U_critical}\textbf{Finite-size scaling for the quantum phase transition between the TBOW and the qSF phase:} (a) Rescaling of the staggered magnetization $m$ as a function of the interaction strength $U/t$ for $\Delta/t = 0.80$ and for different system sizes. The former serves as an order parameter to distinguish between the TBOW phase, where it has a non-zero value, and the qSF phase, where it vanishes. The critical point, $U_c = 7.67t$, is located at the crossing point between the different lines, by assuming the critical exponents of the Ising universality class, $\beta = 1 / 8$ and $\nu = 1$. Inset: These coefficients lead to the collapse of the data to a single line. (b) The location of the critical point is confirmed using the fidelity susceptibility $\chi_F$. This quantity develops a peak near the critical point, and its hight diverges with the system size. In the inset, the critical point is found by extrapolating the location of the peaks for different sizes.}
\end{figure}

We now discuss the the analysis required to calculate one the critical points. In particular, Figure~\ref{fig:U_critical}{\bf (a)} shows the change of $m$ in terms of $U$ for a fixed value of $\Delta$ and for different system sizes. By introducing a proper rescaling factor, we observe how all the lines cross at the quantum critical point $U_c$. In the inset, we show the collapse of the data to a universal line, $m^{\beta/\nu} \sim f\left(L^{1/\nu}(U - U_c)\right)$, where we observe good agreement using the critical exponents of the Ising universality class, $\beta = 1 / 8$ and $\nu = 1$. This contrasts with other transitions in the one-dimensional BHM between an insulator and a SF phase, for which the universality class is of the Kosterlitz-Thouless type \cite{one_dimensional_bosons}. Figure~\ref{fig:U_critical}{\bf (b)} depicts the scaling of the fidelity susceptibility,
\begin{equation}
\chi_F = \lim_{\delta U \to 0} \frac{-2\,\text{log} \, | \langle \psi(U + \delta U) | \psi(U)\rangle |}{\delta U ^2},
\end{equation}
which provides an alternative confirmation of the existence of a quantum phase transition.
This quantity is super-extensive at the critical point \cite{fidelity}, allowing to extract its location  by extrapolating the position of the peak to the thermodynamic limit, $L\rightarrow\infty$ (inset). Since the TBOW phase cannot be adiabatically connected to the non-interacting boson limit ($U = 0$), it can be regarded as an interaction-induced symmetry-broken topological phase, where the interplay between strong correlations and spontaneous symmetry breaking is crucial to stabilize the SPT phase.

\section{Conclusions and Outlook}
\label{sec:conclusions}

In this work, we characterized the TBOW phase that appears in the $\mathbb{Z}_2$BHM at half filling. This one-dimensional model, composed of interacting bosons coupled to a dynamical $\mathbb{Z}_2$ field, shares similarities with fermion-phonon models such as the Su-Schrieffer-Heeger model for polyacetylene. In particular, the $\mathbb{Z}_2$ field can be seen as a simplified version of a dynamical lattice. Focusing first on the hardcore boson limit, we showed how, for a quasi-adiabatic field, the system undergoes a spontaneous breaking of the translational symmetry. This can be regarded as a Peierls transition, where the staggerization of the field opens a gap in the single-particle fermionic spectrum. Using the Zak phase, we characterized this gapped phase as an SPT phase protected by chiral symmetry, where the topological effects coexist with the presence of long-range order. For finite Hubbard  interactions, chiral symmetry is broken, but the phase is still protected by a bond-centered inversion symmetry. Moreover, the spontaneous symmetry breaking remains, even though the standard Peierls mechanism cannot be directly applied in the bosonic case. The TBOW phase extends, therefore, for strong but finite Hubbard interactions. We confirmed numerically our predictions using DMRG. By characterizing the quantum phase transition between the TBOW and a qSF phase for low interactions, we have established the importance of strong correlations to stabilize the former. Our results allow us to regard this phase as an interaction-induced symmetry-breaking topological insulator.

In the future, it would be interesting to extend our numerical results to characterize the full phase diagram of the model, including parameter regimes where the adiabatic approximation is not expected to hold, and to study other BOW phases for different commensurate densities,  characterizing their topological properties using some of the tools hereby presented.

\begin{acknowledgments}
This project has received funding from the European Union's Horizon 2020 research and innovation programme under the Marie Sk\l{}odowska-Curie grant agreement No 665884, the Spanish Ministry MINECO (National Plan 15 Grant: FISICATEAMO No. FIS2016-79508-P, SEVERO OCHOA No. SEV-2015-0522, FPI), European Social Fund, Fundaci\'{o} Cellex, Generalitat de Catalunya (AGAUR Grant No. 2017 SGR 1341 and CERCA/Program), ERC AdG OSYRIS, EU FETPRO QUIC, and the National Science Centre, PolandSymfonia
Grant No. 2016/20/W/ST4/00314. A. D. is financed by a Juan de la Cierva fellowship (IJCI-2017-33180). P. W. acknowledges financial support from the Foundation for Polish Science within the Homing progamme co-financed by the European Union under the European Regional Development Fund. A.B. acknowledges support from the Ram\'on y Cajal program under RYC-2016-20066, spanish MINECO project FIS2015-70856-P, and CAM PRICYT project QUITEMAD+ S2013/ICE-2801.
\end{acknowledgments}

\appendix

\section{Born-Oppenheimer variational approach}
\label{sec:app:details}

In this Appendix, we present various details of the Born-Oppenheimer-type variational ansatz for the ground-state and low-energy excitations of the  $\mathbb{Z}_2$BHM.

For the groundstate, the family of variational states is defined in Eq.~\eqref{eq:BO_var_ansatz}.
The  set of  variational parameters $\{d_{\boldsymbol{n}},\boldsymbol{\theta}\}$ can be fully determined by the minimization of
\beq
\label{eq:e_var_ansatz}
\epsilon_{\rm gs}(\{d_{\boldsymbol{n}},\bs\theta\})=\frac{1}{L}\bra{\Psi_{\rm gs}(\{d_{\boldsymbol{n}},\bs\theta\})} H_{\mathsf{Z_2BH}}^{U\to\infty}\ket{\Psi_{\rm gs}(\{d_{\boldsymbol{n}},\bs\theta\})}.
\eeq
Since the $\mathbb{Z}_2$ fields are quasi-static with respect to bosons, performing average over the former in 
\eqref{eq:e_var_ansatz} leads to the effective Hamiltonian $H_{\rm f}(\boldsymbol{\theta})$ acting on the state of the latter, with $\ket{\psi_{\rm f}(\{d_{\boldsymbol{n}}\}}$ as its ground state. Accordingly, the fermionic variational parameters are fully determined by the $\mathbb{Z}_2$-field variational angles $d_{\boldsymbol{n}}=d_{\boldsymbol{n}}(\boldsymbol{\theta})$ (i.e. the  fermions adapt instantaneously to the slow spins).
As discussed in the main text, for periodic boundary conditions, it suffices to consider only two variational angles, namely $\bs\theta=(\theta_A,\theta_B)$ for the links joining odd-even (even-odd) lattice sites. 
In such a case, the effective Hamiltonian  $H_{\rm f}(\bs\theta)$ turns out to be $\mathsf{BDI}$ (H\"{u}ckel) Hamiltonian $H_{\mathsf{BDI}}(t(\bs\theta),\delta(\bs\theta) )$ of Eq.~\eqref{eq:huckel} parametrized by the variational fields 
\beq
\label{eq:parameters_gs_var2}
\begin{split}
t(\bs\theta)&=t+\frac{\alpha}{2}\big(\sin\theta_A+\sin\theta_B\big),\\
\delta(\bs\theta)&=\frac{\alpha(\sin\theta_A-\sin\theta_B)}{2t+\alpha(\sin\theta_A+\sin\theta_B)}.
\end{split}
\eeq

For such Hamiltonian, the variational ground-state energy \eqref{eq:e_var_ansatz} takes analytical form. To set the notation, and introduce concepts that are also used for the variational ansatz of excitations, 
let us present diagonalisation of  $H_{\mathsf{BDI}}(t(\bs\theta),\delta(\bs\theta) )$. We define Bogoliubov-type fermionic operators as
\beq
\label{eq:A_quasiparticles}
\begin{split}
\gamma_{k,+}=u_k(\bs\theta)c_k+v_k(\bs\theta)c_{k+\pi},\\
\gamma_{k,-}=v_k(\bs\theta)c_k+u_k(\bs\theta)c_{k+\pi}.
\end{split}
\eeq 
where we have used the Fourier transformed operators $c_k=\sum_i\ee^{-\ii ki}c_i/\sqrt{L},$ and $c_{k+\pi}=\sum_i(-1)^i\ee^{-\ii ki}c_i/\sqrt{L}$, the quasi-momenta $k\in[-\frac{\pi}{2},\frac{\pi}{2})$. By using the following functions
\beq
\label{eq:bog_constants}
\begin{split}
u_k(\bs\theta)&=\frac{\ii\,s}{\sqrt{2}}\sqrt{1-\frac{2t(\bs\theta) \cos k}{\epsilon^{\rm f}_k(\bs\theta)}},\\
v_k(\bs\theta)&=\frac{\phantom{\ii}1\phantom{\ii}}{\sqrt{2}}\sqrt{1+\frac{2t(\bs\theta) \cos k}{\epsilon^{\rm f}_{k}(\bs\theta)}},
\end{split}
\eeq
where $s= {\rm sgn}\{\delta (\bs\theta)k\}$ and 
\beq
\label{eq:single_part_energies}
\epsilon^{\rm f}_{k}(\bs\theta)=2t(\bs\theta)\sqrt{\cos^2k+\delta^2(\bs\theta)\sin^2k},
\eeq
one can  rewrite the Hamiltonian $H_{\mathsf{BDI}}(t(\bs\theta),\delta(\bs\theta) )$ in terms of these $\gamma_{k,+}$ and $\gamma_{k,-}$ operators   in  diagonal form
\beq
H_{\mathsf{BDI}}(t(\bs\theta),\delta(\bs\theta) )=\sum_k\epsilon_k^{\rm f}(\boldsymbol{\theta})\left(\gamma_{k,+}^\dagger\gamma_{k,+}^{\phantom{\dagger}}-\gamma_{k,-}^\dagger\gamma_{k,-}^{\phantom{\dagger}}\right),
\eeq
Accordingly, the fermionic part of the variational groundstate $\ket{\psi_{\rm f}(\{d_{\boldsymbol{n}}\})}=\sum_{\boldsymbol{n}}d_{\boldsymbol{n}}(\bs\theta^\star)\ket{\boldsymbol{n}}_{\rm f}$ as follows
\beq
\label{eq:fermGS}
\ket{\psi_{\rm f}(\{d_{\boldsymbol{n}}\})}=\prod_{|k|\leq\pi/2}\gamma^\dagger_{k,-}\ket{0},
\eeq
 where $\ket{0}$ is the fermionic vaccum, and the variational constants $d_{\boldsymbol{n}}=d_{\boldsymbol{n}}(\boldsymbol{\theta})$ only depend on the spin variational angles via Eqs.~\eqref{eq:bog_constants}. As advanced below, the fermions adapt instantaneoulsy to the background $\mathbb{Z}_2$ fields, and the variational angles can be found by minimizing the groundstate energy of Eq.~\eqref{eq:var_energy}, where the first term stems from the addition of the fermionic single-particle energies in Eq.~\eqref{eq:single_part_energies}, while the remaining terms are straightforward expectation values over the spin coherent states.

Let us now turn into the variational ansatz for the low-energy excitations introduced in Eq.~\eqref{eq:BO_var_ansatz_exc}. The excitation energies are then derived from the minimization of
\beq
\label{eq:minimization}
\epsilon_{\rm exc}(\bs \theta^\star)={\rm min}_{\bs \eta}\left(\mathcal{E}[\bs \eta]/\mathcal{N}[\bs \eta]\right),
\eeq
 where 
we have introduced the  norm functional $\mathcal{N}[\bs\eta]=\langle{\Psi_{\rm exc}(\bs\eta)}|{{\Psi_{\rm exc}(\bs \eta)}}\rangle$ and the excitation energy functional
$
\mathcal{E}[\bs\eta]=\bra{\Psi_{\rm exc}(\bs \eta)}H_{\mathsf{Z_2BH}}^{U\to\infty}-\epsilon_{\rm gs}(\bs\theta^\star){\ket{\Psi_{\rm exc}(\bs\eta)}}$. In this part, the Hamiltonian~\eqref{eq:z2_bhm} is treated within the  spin-wave approximation~\eqref{eq:spin-wave} for the $\mathbb{Z}_2$ fields up to quadratic order, such that
\beq
\label{eq:exc_fucntionals}
\begin{split}
\mathcal{N}[\bs\eta]&=\sum_k\eta_{{\rm f},k}^*\eta_{{\rm f},k}^{\phantom{*}}+\sum_i\eta_{{\rm s},i}^*\eta_{{\rm s},i}^{\phantom{*}},\\
\mathcal{E}[\bs\eta]&=\sum_k\epsilon^{\rm f}_{k}(\bs\theta^\star)\eta_{{\rm f},k}^*\eta_{{\rm f},k}^{\phantom{*}}+\sum_i\epsilon^{\rm s}_{i}(\bs\theta^\star)\eta_{{\rm s},i}^*\eta_{{\rm s},i}^{\phantom{*}}.\\
\end{split}
\eeq

By solving $\partial_{\bs\eta^*}(\mathcal{E}[\bs\eta]/\mathcal{N}[\bs\eta])=0$ using $\epsilon_{\rm exc}(\bs \theta^\star)={\rm min}_{\bs\eta}\{\mathcal{E}[\bs\eta]/\mathcal{N}[\bs\eta]\}$, one can see that in the hardcore boson limit, the low-energy excitations can be: {\it (i)} delocalized fermion-like excitations with $\epsilon_{\rm exc}(\bs \theta^\star)=\epsilon^{\rm f}_{k}(\bs \theta^\star)$ $\forall k\in[-\frac{\pi}{2},\frac{\pi}{2})$, or {\it (ii)} localized spin-wave-type excitations with $\epsilon_{\rm exc}(\bs \theta^\star)=\epsilon^{\rm s}_{i}(\bs \theta^\star)$ $\forall i\in\{1,\cdots N\} $, as discussed in the main text.

Let us now give some details on how the calculation can be generalized for softcore bosons~\eqref{eq:perturbation}. Considering the leading-order corrections to the ground-state energy $\epsilon_{\rm gs}(\bs \theta^\star)+\delta\epsilon_{\rm gs}(\bs \theta^\star)$~\eqref{eq:var_energy}, where $\delta\epsilon_{\rm gs}(\bs \theta^\star)=\bra{\Psi_{\rm gs}(\bs \theta^\star)}\Delta H\ket{\Psi_{\rm gs}(\bs \theta^\star)}$, the  excitation energy~\eqref{eq:minimization} will be given by $\epsilon_{\rm exc}(\bs \theta^\star)+\delta\epsilon_{\rm exc}(\bs \theta^\star)$, where
\beq
\delta\epsilon_{\rm exc}(\bs \theta^\star)={\rm min}_{\{\eta\}}\left(\delta\mathcal{{E}}[\eta]/\mathcal{N}[\eta]\right),
\eeq
 and where  the  we have introduced the functional
$
\delta\mathcal{{E}}(\boldsymbol{\eta})=\bra{\Psi_{\rm exc}(\boldsymbol{\eta})}\Delta H-\delta\epsilon_{\rm gs}(\bs \theta^\star){\ket{\Psi_{\rm exc}(\boldsymbol{\eta})}}
$
in terms of the excited-state ansatz~\eqref{eq:BO_var_ansatz_exc}. We can evaluate these corrections by  applying Wick's theorem, as the variational ansatz is built with free spinless fermions. 
Several of the possible Wick contractions will be canceled by the substraction of the ground-state energy shift $\delta\epsilon_{\rm gs}(\bs \theta^\star)$. For the evaluation of the energy gap protecting the TBOW phase,  the non-vanishing contributions will arise from Wick contractions that include single-particle correlations between the excited fermion and the lattice operators, e.g. $\langle \gamma^{\vphantom{\dagger}}_{k,+}c_i^\dagger\rangle_{\rm gs} \langle c_i^{\vphantom{\dagger}} c_{i+1}^{{\dagger}}\rangle_{\rm gs} \langle  c_i^{\vphantom{\dagger}}\gamma_{k,+}^{{\dagger}}\rangle_{\rm gs}$. By performing the  corresponding calculations in detail, we find the particular correction to the energy gap expressed in Eq.~\eqref{eq:gap_closing}.

\bibliographystyle{apsrev4-1}
\bibliography{bibliography}

\end{document}